# The use of multilayer network analysis in animal behaviour

Kelly R. Finn[1], Matthew J. Silk[2], Mason A. Porter[3], and Noa Pinter-Wollman[4]

1. Animal Behavior Graduate Group, University of California, Davis, USA
2. Environment and Sustainability Institute, University of Exeter, UK
3. Department of Mathematics, University of California, Los Angeles, USA
4. Department of Ecology and Evolutionary Biology, University of California, Los Angeles, USA

**Abstract**

Network analysis has driven key developments in research on animal behaviour by providing quantitative methods to study the social structures of animal groups and populations. A recent formalism, known as *multilayer network analysis*, has advanced the study of multifaceted networked systems in many disciplines. It offers novel ways to study and quantify animal behaviour as connected 'layers' of interactions. In this article, we review common questions in animal behaviour that can be studied using a multilayer approach, and we link these questions to specific analyses. We outline the types of behavioural data and questions that may be suitable to study using multilayer network analysis. We detail several multilayer methods, which can provide new insights into questions about animal sociality at individual, group, population, and evolutionary levels of organisation. We give examples for how to implement multilayer methods to demonstrate how taking a multilayer approach can alter inferences about social structure and the positions of individuals within such a structure. Finally, we discuss caveats to undertaking multilayer network analysis in the study of animal social networks, and we call attention to methodological challenges for the application of these approaches. Our aim is to instigate the study of new questions about animal sociality using the new toolbox of multilayer network analysis.

# 1. Introduction

## 1.1 'Multi-dimensionality' of animal social behaviour

Sociality is widespread in animals, and it has a pervasive impact on behavioural, evolutionary, and ecological processes, such as social learning and disease spread (Allen, Weinrich, Hoppitt, & Rendell, 2013; Aplin et al., 2014; Silk, Alberts, & Altmann, 2003; White, Forester, & Craft, 2017). The structure and dynamics of animal societies emerge from interactions between and among individuals (Hinde, 1976; Krause, Croft, & James, 2007; Pinter-Wollman et al., 2014). These interactions are typically 'multi-dimensional', as they occur across different social contexts (e.g., affiliation, agonistic, and feeding), connect different types of individuals (e.g., male–male, female–female, or male–female interactions), and/or vary spatially and temporally. Considering such multi-dimensionality is crucial for thoroughly understanding the structure of animal social systems (Barrett, Henzi, & Lusseau, 2012).

Network approaches for studying the social behaviour of animals have been instrumental in quantifying how sociality influences ecological and evolutionary processes (Krause et al., 2007; Krause, James, Franks, & Croft, 2015; Kurvers et al., Krause, Croft, Wilson, & Wolf, 2014; Pinter-Wollman et al., 2014; Sih, Hanser, & McHugh, 2009; Sueur, Jacobs, Amblard, Petit, & King, 2011; Webber & Vander Wal, 2018; Wey, Blumstein, Shen, & Jordán, 2008). In animal social networks, nodes (also called 'vertices') typically represent individual animals; and edges (also called 'links' or 'ties') often represent pairwise interactions (i.e., behaviours, such as grooming, in which two individuals engage) or associations (e.g., spatio-temporal proximity or shared group memberships) between these individuals. Such a network representation is a simplified depiction of a much more intricate, multifaceted system. A social system can include different types of interactions, with different biological meanings (e.g., cooperative or competitive), which standard network approaches often do not take into account, or they do so by analysing networks of different edge types separately (Gazda, Iyer, Killingback, Connor, & Brault, 2015b). Typical approaches ignore interdependencies that may exist between different types of interactions and between different subsystems (Barrett et al., 2012; Beisner, Jin, Fushing, & McCowan, 2015). Furthermore, networks are often studied as snapshots or aggregations of processes that change over time, but dynamics can play a major role in animal behaviour (Blonder, Wey, Dornhaus, James, & Sih, 2012; Farine, 2018; Wey et al., 2008; Wilson et al., 2014). As we highlighted recently (Silk, Finn, Porter, & Pinter-Wollman, 2018), advances in multilayer network analysis provide opportunities to analyse the multifaceted nature of animal behaviour, to ask questions about links between social dynamics across biological scales, and to provide new views on broad ecological and evolutionary processes. In this paper, we introduce the new mathematical formalism of multilayer network analysis to researchers in animal behavior. This formalism provides a common vocabulary to describe, compare, and contrast multilayer network methodologies. Our goal is to review research areas and questions in animal behavior that are amenable to multilayer network analysis, and we link specific analyses to these questions (see Table 1). In the remainder of this section, we describe different types of multilayer networks and detail how they can encode animal data. In Section 2, we review several questions and hypotheses, across social scales, that multilayer network analysis can help investigate. We summarize key questions and provide a guide to available methods and software for multilayer network analysis in Table 1. Throughout Section 2, we present worked examples to illustrate our ideas. In Section 3, we consider some of the requirements and caveats of multilayer network analysis as a tool to study animal social behaviour and discuss several directions for future work.

## 1.2 What are multilayer networks?

Multilayer networks are assemblages of distinct network 'layers' that are connected (and hence coupled) to each other via interlayer edges (Boccaletti et al., 2014; Kivelä et al., 2014). A multilayer network can include more than one 'stack' of layers, and each such facet of layering is called an 'aspect'. For instance, one aspect of a multilayer network can encode temporal dynamics and another aspect can represent the types of social interactions (Fig. 1 and Appendix I).

The recent formalism of *multilayer networks* has opened up new ways to study multifaceted networked systems (Boccaletti et al., 2014; Kivelä et al., 2014). The application of multilayer networks to questions in animal behaviour is still in its infancy, but multilayer network analysis has facilitated substantial advances over monolayer (i.e., single-layer) network analysis in many other fields (Aleta & Moreno (2018) and Kivelä et al. (2014)). For example, multilayer network approaches have made it possible to identify important nodes that are not considered central in a monolayer network (De Domenico, Solé-Ribalta, Omodei, Gómez, & Arenas, 2015). Multilayer approaches applied to studying information spread on Twitter (where, e.g., one can use different layers to represent 'tweets', 'retweets', and 'mentions') have uncovered information spreaders who have a disproportionate impact on social groups but were overlooked in prior monolayer investigations (Al-Garadi, Varathan, Ravana, Ahmed, & Chang, 2016). Multilayer modelling of transportation systems has improved investigations of congestion and efficiency of transportation. For example, each layer may be a different airline (De Domenico, Solé-Ribalta, et al., 2015) or a different form of transportation in a city (Chodrow et al., 2016; Gallotti & Barthélemy, 2015; Strano, Shai, Dobson, & Barthélemy, 2015). Modelling dynamical processes on multilayer networks can result in qualitatively different outcomes compared to modelling dynamics on aggregate representations of networks (see Appendix II for a discussion of aggregating networks) or on snapshots of networks (De Domenico, Granell, Porter, & Arenas, 2016). For instance, the dynamics of disease and information spread can be coupled in a multilayer framework to reveal how different social processes can impact the onset of epidemics (Wang, Andrews, Wu, Wang, & Bauch, 2015). Historically, the usage of 'multiplexity' dates back many decades (Mitchell, 1969), and the new mathematical formalism (De Domenico et al., 2014; Kivelä et al., 2014; Newman, 2018c; Porter, 2018) has produced a unified framework that makes it possible to consolidate analysis and terminology. For reviews of previous multilayer network studies and applications in other fields, see (Aleta & Moreno, 2018; Boccaletti et al., 2014; D'Agostino & Scala, 2014; Kivelä et al., 2014; Pilosof, Porter, Pascual, & Kéfi, 2017).

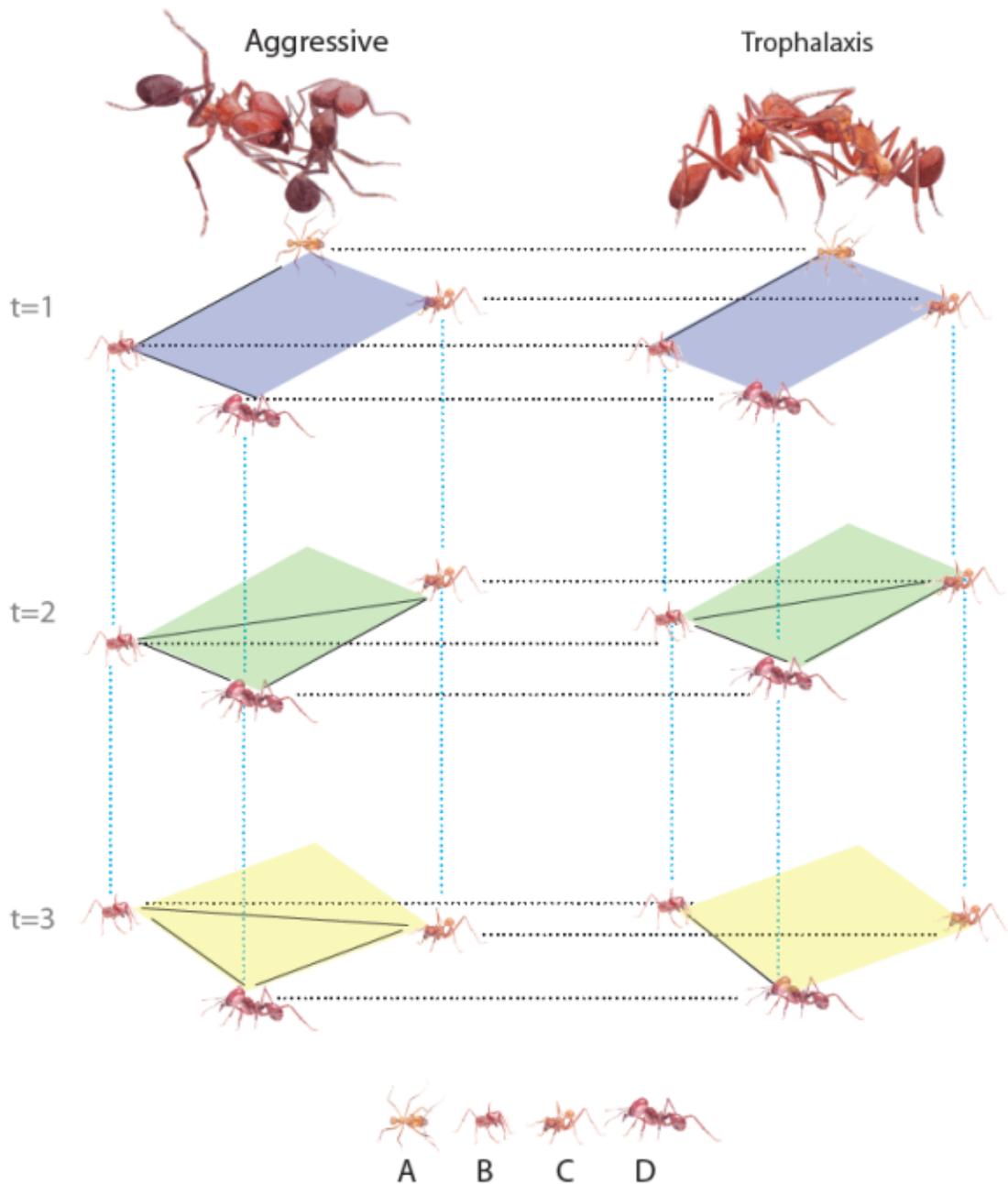

**Figure 1: A hypothetical multilayer network.** Four ants interact at different time points and in two different ways. Each diamond represents a layer. The stack of three layers on the left represents aggressive interactions, and the stack of three layers on the right represents trophalactic interactions. Each colour represents a different time point (blue is *t*=1, green is *t*=2, and yellow is *t*=3). Solid lines represent intralayer (i.e., within-layer) interactions, dotted blue

lines represent interlayer (i.e., across-layer) relationships in the temporal aspect, and dotted black lines represent interlayer edges in the behavioural aspect. These interlayer interactions connect replicates of the same individuals across different layers. See Appendix I for further discussion and for a presentation of the mathematical formalism.

**1.3 Types of multilayer networks**

The mathematical framework of multilayer networks was developed recently to create a unified formalism (De Domenico et al., 2014; Kivelä et al., 2014; Mucha, Richardson, Macon, Porter, & Onnela, 2010; Porter, 2018). One can use this multilayer network framework, which we follow in this paper and detail in Appendix I, to represent a variety of network types and situations. In contrast to monolayer networks, which are traditional in network analysis and which consist of only a single network 'layer', multilayer networks can include many different types of data that are commonly collected in studies of animal behaviour. For example, types of social interactions, spatial locations (with connections between them), and different measures of genetic relatedness can all constitute layers in a multilayer network. Node attributes can include behavioural or physical phenotypes, sex, age, personality, and more. Edge attributes, such as their weight or direction, can encode interaction frequencies, distances between locations, dominance, and so on. Commonly studied variants of multilayer networks that can accommodate such data include the following.

(1) ***Multiplex networks*** (i.e., ***edge-coloured networks***) are networks in which interlayer edges connect nodes to themselves on different layers (Fig. 1 and Appendix I). It is often assumed, for convenience, that all layers consist of the same set of nodes, but this is not necessary.

a. In ***multirelational networks,*** each layer represents a different type of interaction. For example, a network of aggressive interactions can be connected with a network of affiliative interactions through interlayer edges that link individuals to themselves if they appear in both layers (Fig. 1; horizontal dotted black lines).

b. In ***temporal networks,*** each layer encodes the same type of interactions during different time points or over different time windows. In the most common multiplex representation of a temporal network, consecutive layers are connected to each other through interlayer edges that link individuals to themselves at different times (Fig. 1; vertical dotted blue lines).

(2) In ***interconnected networks*** (i.e., ***node-coloured networks***)*,* the nodes in different layers do not necessarily represent the same entities, and interlayer edges can exist between different types of nodes. (See our discussion of the mathematical formalism and an example figure in Appendix I.)

a. ***Networks of networks*** consist of subsystems, which themselves are networks that are linked to each other through interlayer edges between the subsystems' nodes. For example, one can model intra-group interactions in a population-level network of interactions between social groups, which are themselves networks.

b. In ***inter-contextual networks,*** one can construe each layer as representing a different type of node. For example, interactions between males can be in one layer, interactions between females can be in a second layer, and inter-sex

> interactions are interlayer edges. See Fig. 1 of Silk, Weber, et al. (2018) and Fig 1. of Silk, Finn, Porter, & Pinter-Wollman (2018).
>
> c. ***Spatial networks***, which we define here as networks of locations, can be linked with social networks of animals that move between these locations (Pilosof et al., 2017; Silk, Finn, Porter, & Pinter-Wollman, 2018). Our use of the term "spatial networks" refers to networks that are embedded in space, rather than networks that are influenced by a latent space (Barthélemy, 2018).

Throughout this paper, we use the term "multilayer networks" to refer to any of the variants above, unless we specify that a method applies to only one or a subset of specific network types. For a review of other types of multilayer networks, see (Kivelä et al., 2014).

**2. Novel insights into animal sociality: From individuals to populations**

We propose that a multilayer network approach can advance the study of animal behaviour and expand the types of questions that one can investigate. Specifically, we discuss how a multilayer framework can enhance understanding of (1) an individual's role in a social network, (2) group-level structure and dynamics, (3) population structure, and (4) evolutionary models of the emergence of sociality.

**2.1 An individual's role in society**

Traditionally, the use of network analysis to examine the impact of individuals on their society has focused on the social positions of particular individuals using various centrality measures (such as degree, eigenvector centrality, betweenness centrality, and others (Pinter-Wollman et al., 2014; Wasserman & Faust, 1994; Wey et al., 2008; Williams & Lusseau, 2006)).

It is common to construe individuals with disproportionally large centrality values as influential or important to a network in some way (but see (Rosenthal, Twomey, Hartnett, Wu, & Couzin, 2015) for a different trend). The biological meaning of 'importance' and corresponding centrality measures differ among types of networks and is both system-dependent and question-specific. Consequently, one has to be careful to avoid misinterpreting the results of centrality calculations. Centrality measures have been used to examine which individuals have the most influence on a group in relation to age, sex, or personality (Sih et al., 2009; Wilson, Krause, Dingemanse, & Krause, 2013) and to study the fitness consequences of holding an influential position (Pinter-Wollman et al., 2014). A multilayer approach can advance understanding of roles that individuals play in a population or a social group, and it can potentially identify central individuals who may be overlooked when using monolayer approaches on "multidimensional" data.

An individual's role in a social group is not restricted to its behaviour in just one social or ecological situation. A multilayer approach creates an opportunity to consolidate analyses of a variety of social situations and simultaneously examine the importances of individuals across and within situations. Many centrality measures have been developed for multilayer networks, and different ones encompass different biological interpretations. For instance, eigenvector 'versatility' (see Appendix I for its mathematical definition) is one way to quantify the overall importance of individuals across and within layers, because this measure takes into account multiple layers to identify individuals who increase group cohesion in multiple layers and bridge social situations (De Domenico, Solé-Ribalta, et al., 2015). In a multirelational network, an individual can have small degree (i.e., degree centrality) in each layer, which each represent a different social situation, but it may participate in many social situations, thereby potentially producing a larger impact on social dynamics than individuals with large degrees in just one or a few social situations. One can also account for the interrelatedness of behaviours in different layers in a multilayer network when combining interlayer centralities, if appropriate for the study system (De Domenico, Solé-Ribalta, et al., 2015). For example, it is not possible for two individuals to engage in grooming interactions without also being in proximity. By accounting for interrelatedness between proximity and grooming when calculating multilayer centralities and versatilities, it may be possible to consider grooming interactions as explicitly constrained by proximity interactions and thereby incorporate potentially important nuances.

The appropriateness of a versatility measure differs across biological questions, just as various centrality measures on a monolayer network have different interpretations (Wasserman & Faust, 1994; Wey et al., 2008). Versatility measures that have been developed include shortest-path betweenness versatility, hub/authority versatility, Katz versatility, and PageRank versatility (De Domenico, Solé-Ribalta, et al., 2015). Experimental removal of versatile nodes, similar to the removal of central nodes in monolayer networks (Barrett et al., 2012; Firth et al., 2017; Flack et al., 2006; Pruitt & Pinter-Wollman, 2015; Sumana & Sona, 2013), has the potential to uncover the effects of the removed nodes on group actions, group stability, and their impact on the social positions of other individuals. However, which versatility measure gives the most useful information about an individual's importance may depend on the level of participation of an individual in the different types of behaviours that are encoded in a multilayer network. Further, if layers have drastically dissimilar densities, one layer can easily dominate a versatility measure (Braun et al. 2018); see our discussion of caveats in Section 3. In addition to calculating node versatility, one can examine versatility of edges to yield interesting insights into the importances of relationships with respect to group stability and cohesion. Such an approach can help reveal whether interlayer interactions are more important, less important, or comparably important than intralayer interactions for group cohesion. Examining edge versatility can also illuminate which interactions between particular individuals (within or across layers) have the largest impact on group activity and/or stability; and it may be helpful for conservation efforts, such as in identifying social groups that are vulnerable to fragmentation (Snijders, Blumstein, Stanley, & Franks, 2017).

A multilayer approach can help elucidate the relative importances of different individuals in various social or ecological situations. For example, a node's 'multidegree' is a vector of the intralayer degrees (each calculated as on a monolayer network) of an individual in each layer. Differences in how the degrees of individuals are distributed across layers help indicate which individuals have influence over others in the different layers. For example, if each layer represents a different situation, individuals whose intralayer degree peaks in one situation may be more influential in that context than individuals whose intralayer degree is small in that situation but peaks in another one. Because multidegree does not account for interlayer connections, quantitatively comparing it with versatility or other multilayer centralities, which account explicitly for interlayer edges (Kivelä et al., 2014), can help elucidate the importance of interlayer edges and thereby highlight interdependencies between biological situations. Such behavioural interdependencies may elucidate and quantify the amount of behavioural carryover across situations (i.e., 'behavioural syndromes') (Sih, Bell, & Johnson, 2004) if, for example, measures that account for interlayer edges explain observed data better than measures that do not take into account such interdependencies.

As a final example, one can use a multilayer approach to examine temporal changes in an individual's role (or roles) in a group. A multilayer network in which one aspect represents time and another aspect represents situation (Fig. 1) can reveal when individuals gain or lose central roles and whether roles are lost simultaneously in all situations or if changes in one situation precede changes in another. Comparing monolayer (e.g., time-aggregated) measures and multilayer measures has the potential to uncover the importance of temporal changes in an animal's fitness.

**2.1.1 Roles of individuals in a group: Baboon versatility in a multiplex affiliation network**

To demonstrate the potential insights from employing multilayer network analysis to examine the roles of individuals in a social group using multiple interaction types, we analysed published affiliative interactions from a baboon *Papio cynocephalus* group of 26 individuals (Franz, Altmann, & Alberts, 2015b, 2015a) (Fig. 2). One can quantify affiliative relationships in primates in multiple ways, including grooming, body contact, and proximity (Barrett & Henzi, 2002; Jack, 2003; Kasper & Voelkl, 2009; Pasquaretta et al., 2014). To characterize affiliative relationships, combinations of these behaviours have been investigated separately (Jack, 2003; Perry, Manson, Muniz, Gros-Louis, & Vigilant, 2008), pooled together (Kasper & Voelkl, 2009), or used interchangeably (Pasquaretta et al., 2014). These interaction types are often correlated with each other, but their networks typically do not coincide completely (Barrett & Henzi, 2002; Brent, MacLarnon, Platt, & Semple, 2013).

We analyse the baboon social data in four ways: (1) as a weighted grooming network with only grooming interactions (Fig. 2A), (2) as a weighted association network with only proximity-based associations (Fig. 2B), (3) as an aggregate monolayer network that we obtained by summing the weights of grooming and association interactions of the node pairs (Fig. 2C; see Appendix II for more details on aggregating networks), and (4) as a multiplex network with two layers (one for grooming and one for associations). We then calculated measures of centrality (for the monolayer networks in (1)–(3)) and versatility (for the multilayer network (4)) using MuxViz (De Domenico et al. 2015). We ranked individuals according to their PageRank centralities and versatilities (De Domenico, Solé-Ribalta, et al., 2015), which quantify the centrality of an individual in a network recursively based on being adjacent to central neighbours (Fig. 3).

The most versatile baboon in the multilayer network (individual 3 in Fig. 3) was the fourth-most central individual in the aggregated network, the second-most central individual in the grooming network, and 16th-most central individual in the association network (Fig. 3). These differences in results using the multilayer, aggregated, and independent networks of the same data highlight the need to (1) carefully select which behaviours to represent as networks and (2) interpret the ensuing results based on the questions of interest (Carter, DeChurch, Braun, & Contractor, 2015; Krause et al., 2015). When social relationships depend on multiple interaction types, it is helpful to use a multilayer network framework to reliably capture an individual's social roles (see Table 1 for more questions and tools), because monolayer calculations may yield different results and centrality in one layer can differ substantially from versatility in an entire multilayer network (Fig. 3).

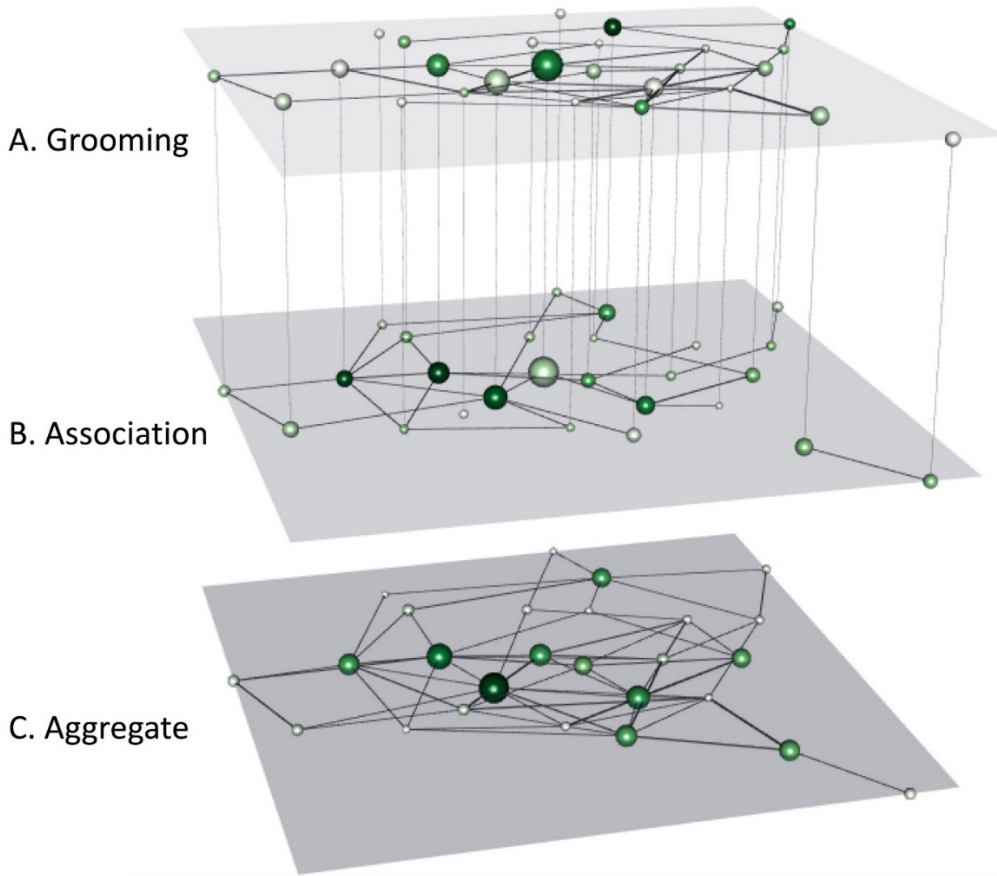

**Figure 2:** Social networks of a baboon group based on (A) grooming interactions, (B) proximity-based association relations, and (C) an aggregate of both interaction types. We created the network visualization using MuxViz (De Domenico, Porter, & Arenas, 2015). To construct a multilayer network, we joined the grooming and association monolayer networks as two layers in a multiplex network by connecting nodes that represent the same individual using interlayer edges. The sizes of the nodes are based on multilayer PageRank versatility (with larger nodes indicating larger versatilities). We colour the nodes based on monolayer PageRank centrality (with dark shades of green indicating larger values). A given individual in these two layers has the same size, but it can have different colours in the two layers. In the aggregate layer, we determine both the node sizes and their colours from PageRank centrality values in the aggregate

network. We position the nodes in the same spatial location in all three layers. The data (Franz et al., 2015a) are from (Franz et al., 2015b).

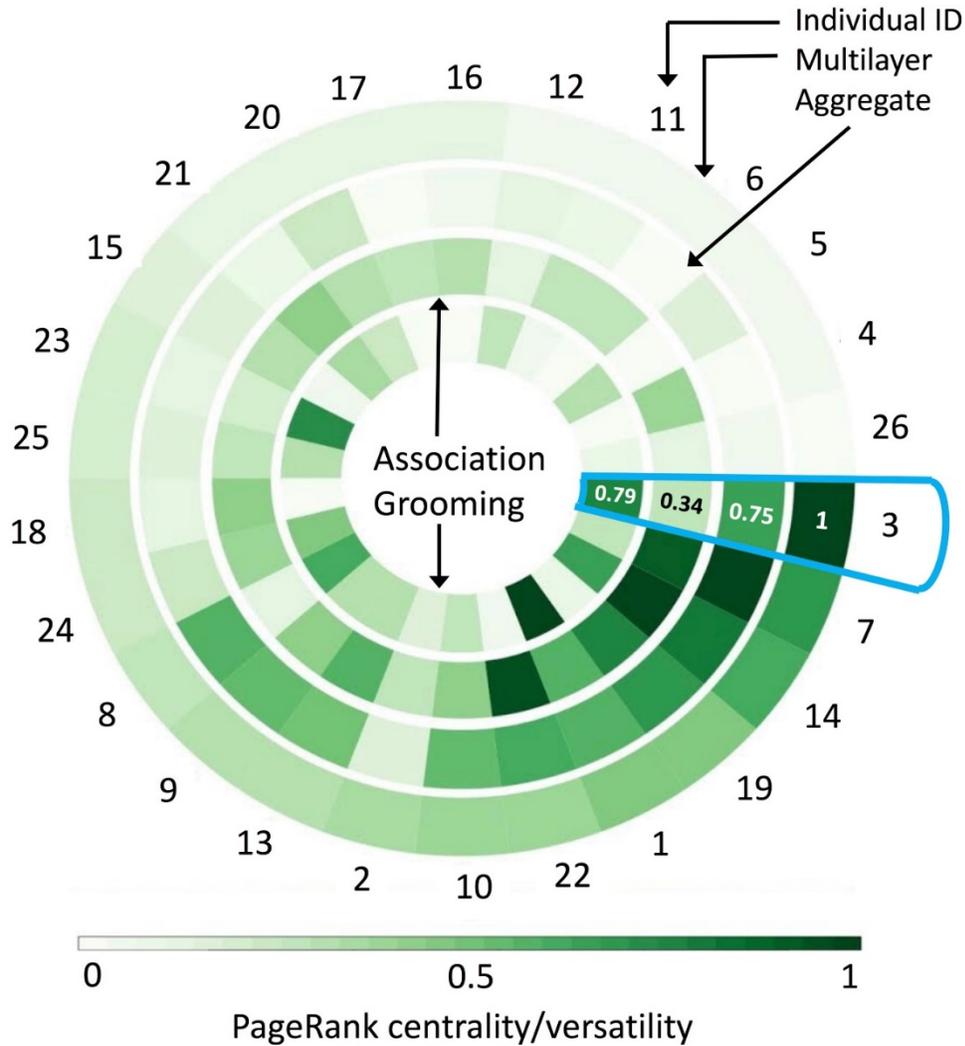

**Figure 3:** A circular heat map illustrates variation among individuals in PageRank centralities and versatilities. Darker colours indicate larger values of PageRank centralities and versatilities. A given angular wedge in the rings indicates the values for one individual, whose ID we list outside the ring. The rings are PageRank centrality values from the monolayer grooming network (innermost ring), association network (second ring), aggregate network in which we sum the grooming and association ties (third ring), and PageRank versatility for the multiplex network

(outermost ring). Using a blue outline, we highlight individual 3, who we discuss in the main text. We indicate the PageRank centrality and versatility values of individual 3 on the rings. We created this visualization using MuxViz (De Domenico, Porter, et al., 2015). The data (Franz et al., 2015a) are from (Franz et al., 2015b).

**2.2 Multilayer structures in animal groups**

Animal social groups are emergent structures that arise from local interactions (Sumpter, 2010), making network analysis an effective approach for examining group-level behaviour. Networks provide useful representations of dominance hierarchies (Hobson et al. 2013) and allow investigations of information transmission efficiency (Pasquaretta et al., 2014), group stability (Baird & Whitehead, 2000; McCowan et al., 2011), species comparisons (Pasquaretta et al., 2014; Rubenstein, Sundaresan, Fischhoff, Tantipathananandh, & Berger-Wolf, 2015), and collective behaviour (Rosenthal et al., 2015; Westley, Berdahl, Torney, & Biro, 2018). However, given that animals interact with each other in many different—and potentially interdependent—ways, a multilayer approach may help accurately capture a group's structure and/or dynamics. In one recent example, Smith-Aguilar et al. (2018) studied a six-layer multiplex network of spider monkeys, with layers based on types of interactions. In this section, we detail how multilayer methodologies can advance the study of group stability, group composition, and collective movement.

One can analyse changes in group stability and composition using various multilayer calculations or by examining changes in relationships across network layers (Beisner & McCowan, 2015). For instance, Barrett et al. (2012) examined changes in a baboon group following the loss of group members by calculating a measure from information theory called

'joint entropy' on a multiplex network—with grooming, proximity, and aggression layers—both before and after a known perturbation. A decrease in joint entropy following individual deaths corresponded to individuals interacting in a more constrained and therefore more predictable manner. Using a different approach, Beisner et al. (2015) investigated co-occurrences of directed aggression and status-signalling interactions between individuals in macaque behavioural networks. In their analysis, they employed a null model that incorporates constraints that encode interdependences between behaviour types. For example, perhaps there is an increased likelihood that animal *B* signals to animal *A* if animal *A* aggresses animal *B*. These constraints were more effective at reproducing the joint probabilities (which they inferred from observations) of interactions in empirical data in stable macaque groups than in groups that were unstable and eventually collapsed (Chan, Fushing, Beisner, & McCowan, 2013). These findings illustrate how interdependencies between aggression and status-signalling network layers may be important for maintaining social stability in captive macaque groups. A potential implication of this finding is that analysing status-signalling and aggression may be helpful for predicting social stability. Another approach that may be useful for uncovering temporal structures in multilayer networks is an extension of stochastic actor-oriented models (SAOMs) (Snijders, 2017). One can use SAOMs to examine traits and processes that influence changes in network ties over time, including in animal social networks (Fisher, Ilany, Silk, & Tregenza, 2017; Ilany, Booms, & Holekamp, 2015). SAOMs can use unweighted or weighted edges, with some restrictions in how weights are incorporated (Snijders, 2017). A multiple-network extension to an SAOM enables modelling of the co-dynamics of two sets of edges, while incorporating influences of other individual or network-based traits. Such an approach has the potential to provide interesting insights into how changes in one layer may cascade into changes in other layers. It also provides a useful method to quantify links between group-level structural changes and temporal dynamics of individual centralities.

Multilayer analysis of animal groups can go beyond monolayer network analysis when highlighting a group's composition and substructures. For example, one measure of interdependence, the proportion of shortest paths between node pairs that span more than one layer (Morris & Barthélemy, 2012; Nicosia, Bianconi, Latora, & Barthélemy, 2013), can help describe a group's interaction structure. In social insect colonies, layers can represent different tasks. As time progresses and individuals switch tasks, an individual can appear in more than one layer. The amount of overlap among layers (see Section 3 of Appendix I for examples of overlap measures) can indicate the level of task specialization and whether or not there are task-generalist individuals (Pinter-Wollman, Hubler, Holley, Franks, & Dornhaus, 2012). Consequently, the above interdependence measure may be useful as a new and different way to quantify division of labour (Beshers & Fewell, 2001), because having a small proportion of shortest paths that traverse multiple layers may be an indication of pronounced division of labour. Such a new measure may reveal ways in which workers are allocated to tasks that are different from those that have been inferred by using other measures of division of labour. Comparing different types of measures may uncover new insights into the mechanisms that underlie division of labour.

Animal groups are often organized into substructures called `communities' (Fortunato & Hric, 2016; Porter, Onnela, & Mucha, 2009; Shizuka et al., 2014; Wolf, Mawdsley, Trillmich, & James, 2007), which are sets of individuals who interact with each other more frequently (or more often) than they do with other individuals. Finding communities can aid in predicting how a group may split, which can be insightful for managing captive populations when it is necessary to remove individuals (Sueur, Jacobs, et al., 2011). Community-detection algorithms distinguish sets of connected individuals who are more connected within a community than with other communities in a network. One example of a multilayer community-detection algorithm is maximization of 'multislice modularity' (Mucha et al., 2010), which can account for different behaviours and/or time windows. A recent review includes a discussion of how multilayer modularity maximization can inform ecological questions, such as the ecological effects of interdependencies between herbivory and parasitism (Pilosof et al., 2017). In animal groups, individuals can be part of more than one community, depending on the types of interactions under consideration. For example, an individual may groom with one group of animals but fight with a different group. Because maximizing multislice modularity does not constrain an individual's membership to a single community, it can yield communities of different functions with overlapping membership. It can also be used to examine changes in community structure over time. Additionally, sex, age, and kinship are known to influence patterns of subgrouping in primates (Sueur, Jacobs, et al., 2011), so investigating group structure while considering several of these characteristics at once can reveal influences of subgrouping (such as nepotism) that may not be clear when using monolayer clustering approaches. See Aleta & Moreno (2018) for references to various methods for studying multilayer community structure.

Collective motion is another central focus in studies of animal groups (Berdahl, Biro, Westley, & Torney, 2018; Sumpter, 2010). Coordinated group movements emerge from group members following individual-based, local rules (e.g., in fish schools and bird flocks) (Couzin, Krause, James, Ruxton, & Franks, 2002; Sumpter, 2010). Recent studies of collective motion have employed network analysis to examine relationships of individuals beyond the ones with their immediate neighbours. For instance, one can incorporate connections between individuals who are in line of sight of each other (Rosenthal et al., 2015) or with whom there is a social relationship in other contexts (Bode, Wood, & Franks, 2011; Farine et al., 2016). One can also combine multiple sensory modes into a multilayer network to analyse an individual's movement decisions. Expanding the study of collective motion to incorporate multiple sensory modalities (e.g., sight, odour, vibrations, and so on) and social relationships (e.g., affiliative, agonistic, and so on) can benefit from a multilayer network approach, which may uncover synergies among sensory modes, social relationships, and environmental constraints.

**2.2.1 Multilayer groupings: Dolphin communities emerge from multirelational interactions**

To demonstrate the utility of multilayer network analysis for uncovering group dynamics, we analysed the social associations of 102 bottlenose dolphins that were observed by (Gazda et al., 2015b). Gazda et al, (2015b) recorded dolphin associations during travel, socialization, and feeding. They identified different communities when analysing the interactions as three independent networks and compared the results with an aggregated network, in which they treated all types of interactions equally (regardless of whether they occurred when animals were traveling, socializing, or foraging). However, analysing these networks separately or as one aggregated network ignores interdependencies that may exist between the different behaviours (Kivelä et al., 2014). Therefore, we employed multiplex community detection, using the multilayer InfoMap method (De Domenico et al. 2015), to examine how interdependency between layers influences which communities occur when the data are encoded as a multiplex network. We use multiplex community detection to assign each replicate of an individual (there is one for each layer in which an individual appears; Appendix I) to a community. Therefore, an individual can be assigned to one or several communities, where the maximum number corresponds to the number of layers in which the individual is present. The community assignments depend on how individuals are connected with each other in a multilayer network and on interactions between layers, which arise in this case from a parameter in the multilayer InfoMap method (see Appendix II for details). The coupling between layers thus arises both from interlayer edges and their weights (Appendix I) and from a parameter in the community-detection method (Appendix II). With no coupling, the layers are distinct and communities cannot span more than one layer; for progressively larger coupling, communities span multiple layers increasingly often. For details on our parameter choices for community detection with the multilayer InfoMap method, see Appendix II.

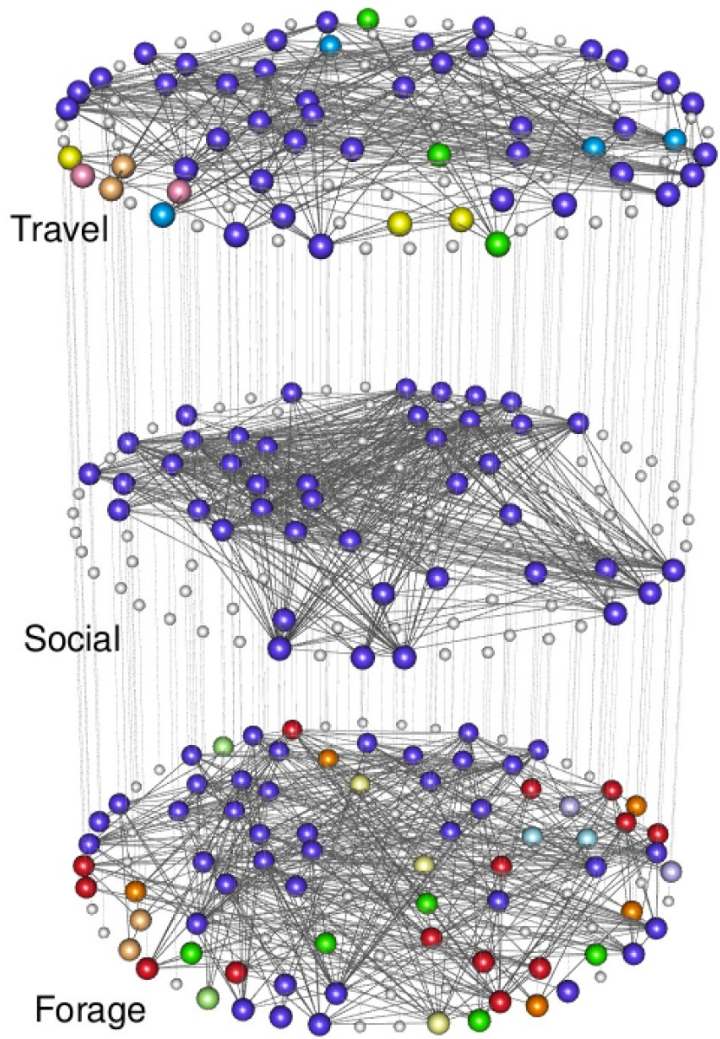

**Figure 4:** Multiplex network of dolphin proximity-based associations during (1) traveling, (2) socializing, and (3) foraging. There are 102 distinct individuals, and each layer has a node for each individual. Individuals who were never seen interacting in a specific layer (behavioural context) are the smaller white nodes. Individuals who interacted in at least one layer are the larger nodes, which we colour based on their community assignment from multilayer InfoMap (De Domenico et al. 2015). We created the network visualization with MuxViz (De Domenico et al. 2015). The data (Gazda, Iyer, Killingback, Connor, & Brault, 2015a) are from (Gazda et al., 2015b).

To be consistent with Gazda et al, (2015b), our multiplex network (Fig. 4) includes only individuals who were seen at least 3 times, and we weight the edges using the half-weight index (HWI) of association strength (Cairns & Schwager, 1987). Our community-detection computation yielded 12 communities. The largest community (dark blue; Fig. 4) consists of individuals from all three association layers, and several smaller communities consist of only foraging individuals, only traveling individuals, and both foraging and traveling individuals. For details on the specific implementation of the InfoMap method, see Appendix II.

In their investigation, Gazda et al, (2015b) revealed contextually-dependent association patterns, as indicated by different numbers of communities in the foraging (17), travel (8), and social (4) networks. Notably, when considering the three behavioural situations as a multiplex network, we found similar trends in the numbers of communities across behavioural situations: foraging individuals are in 9 communities, traveling individuals are in 6 communities, and individuals who interact socially are in only 1 community. Thus, our analysis strengthens the finding that dolphins forage in more numerous, smaller groups and socialize in fewer, larger groups. Different methods for community detection yield different communities of nodes (Fortunato & Hric, 2016) therefore, it is not surprising that we detected a different number of communities in the monolayer networks than the number in Gazda et al, (2015b). We used InfoMap, which has been implemented for both monolayer and multilayer networks. By contrast, Gazda et al, (2015b) used a community-detection approach that has been implemented only for monolayer networks. Additionally, because we find one markedly larger community that spans all layers, it may also be useful to explore core–periphery structure in this network (Csermely, London, Wu, & Uzzi, 2013; Rombach, Porter, Fowler, & Mucha, 2017).

We also analysed each layer independently and an aggregate of all layers using monolayer InfoMap (Rosvall & Bergstrom, 2007), which is implemented in MuxViz. Multiplex community detection produces somewhat different community assignments from monolayer community detection (Fig. 5). With a multiplex network, one can identify and label an individual's membership in a community that spans one or several layer(s) (Fig. 5A). However, in monolayer community detection, one examines individuals independently in different layers, thereby assigning their replicates in different layers to different communities (Fig. 5B). Therefore, which individuals are grouped into communities can vary substantially. (See Table 1 for more questions and tools in multilayer community detection in animal behaviour). As this example illustrates, depending on the research aims, the form of the data, and knowledge of the study system, one or both of monolayer and multilayer investigations may provide valuable insights into the structure of a social system of interest.

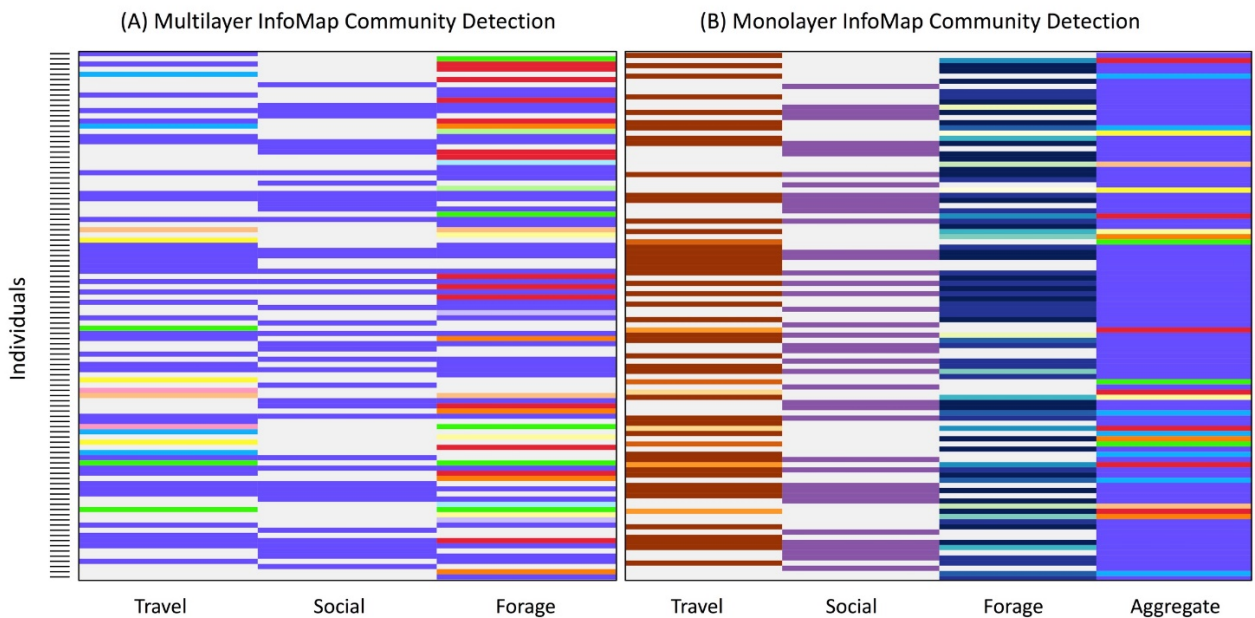

**Figure 5:** Community structures of individuals from (A) a multilayer InfoMap community detection and (B) monolayer InfoMap community detection. Each row represents an individual dolphin, and each column represents a behavioural situation. In the multiplex community detection (A), communities can span all three columns of behaviours, and individuals who are the same colour in one or more columns belong to the same community. Community colours are the same as those that we used in Figure 4. Note that an individual who appears in all three layers can be assigned to the same community in all three situations (and therefore have the same colour in all three columns). It can also be part of three different communities, and it then has different colours in each layer. It can also be assigned twice to one community and once to another. In monolayer InfoMap (B), each behavioural situation (as well as the aggregated monolayer network in the last column) yields a separate set of communities, so we use a different colour palette in each column. Individuals in the same column and the same colour are assigned to the same community. In both panels, white represents individuals who do not exist in the associated behavioural situation.

## 2.3 Multilayer processes at a population level

Network analysis has been fundamental in advancing understanding of social processes over a wide range of spatial scales and across multiple social groups (Silk, Croft, Tregenza, & Bearhop, 2014; Sueur, King, et al., 2011). A multilayer approach is convenient for combining spatial and social networks (e.g., in a recent study of international human migration (Danchev & Porter, 2018)), and it may contribute to improved understanding of fission–fusion dynamics, transmission processes, and dispersal. It also provides an integrative framework to merge social data from multiple species and extend understanding of the drivers that underlie social dynamics of multi-species communities (Farine, Garroway, & Sheldon, 2012; Sridhar, Beauchamp, & Shanker, 2009).

Many animals possess complicated fission–fusion social dynamics, in which groups join one another or split into smaller social units (Couzin & Laidre, 2009; Silk et al., 2014; Sueur, King, et al., 2011). It can be insightful to study such populations as networks of networks. Recent advances in quantifying temporal dynamics of networks have shed some light on fission–fusion social structures (Rubenstein et al., 2015). A multilayer approach applied to association data (collected at times that make it is reasonable to assume that group membership is independent between observations) can assist in detecting events and temporal scales of social transitions in fission-fusion societies. For example, if each layer in a multiplex network represents the social associations of animals at a certain time, a multiplex community-detection algorithm can uncover temporally cohesive groups, similar to the detection of temporal patterns of correlations between various financial assets (Bazzi et al., 2016). Further development of community detection and other clustering methods for general multilayer networks (e.g., stochastic block models (Peixoto, 2014, 2015) and methods based on random walks (De Domenico, Lancichinetti, et al., 2015; Jeub, Balachandran, Porter, Mucha, & Mahoney, 2015; Jeub, Mahoney, Mucha, & Porter, 2017) may provide insights into the social and ecological processes that contribute to the temporal stability of social relationships in fission–fusion societies.

Ecological environments and connections between different locations have fundamental impacts on social dynamics (Firth & Sheldon, 2016; Spiegel, Leu, Sih, & Bull, 2016). A multilayer network representation can explicitly link spatial and social processes in one framework (Pilosof et al., 2017). One approach is to use interconnected networks of social interactions and spatial locations to combine layers that represent social networks with layers for animal movement and habitat connectivity. Data on social interactions can also have multiple layers, with different layers representing interactions in different locations or habitats. For example, in bison *Bison bison,* it was observed that group formation is more likely in open-meadow habitats than in forests (Fortin et al., 2009). The same study also noted that larger groups are more likely than smaller groups to occur in meadow habitats. Multilayer network approaches, such as examining distributions of multilayer diagnostics, may be helpful for detecting fundamental differences in social relationships between habitats.

  Important dynamical processes in animal societies, such as information and disease transmission, are intertwined with social network structures (Allen et al., 2013; Aplin et al., 2014; Aplin, Farine, Morand-Ferron, & Sheldon, 2012; Hirsch, Reynolds, Gehrt, & Craft, 2016; Weber et al., 2013). Research on networks has revealed that considering multilayer network structures can produce very different spreading dynamics than those detected when collapsing (e.g., by aggregating) multiple networks into one monolayer network (De Domenico et al., 2016). Multilayer approaches can uncover different impacts on transmission from different types of social interactions (Craft 2015; White et al. 2017) or link the transmission of multiple types of information or disease across the same network. Compartmental models of disease spreading, which describe transitions of individuals between infective and other states (e.g., susceptible–infected [SI] models, susceptible–infected–recovered [SIR] models, and others) (Kiss, Miller, & Simon, 2017) have been used to model transmission through multilayer networks (Aleta & Moreno, 2018; De Domenico et al., 2016; Kivelä et al., 2014). For example, several studies have incorporated a multilayer network structure into an SIR model for disease spreading coupled with information spreading about the disease, with the two spreading processes occurring on different network layers (Wang, Andrews, Wu, Wang, & Bauch, 2015). This approach suggests that taking into account the spread of information about a disease can reduce the expected outbreak size, especially in strongly modular networks and when infection rates are low (Funk, Gilad, Watkins, & Jansen, 2009). Given the growing evidence for coupled infection and behaviour dynamics in animals (Croft, Edenbrow, et al., 2011; Lopes, Block, & König, 2016; Poirotte et al., 2017), using multilayer network analysis to help understand interactions between information and disease spread is likely to be informative in studies of animal contagions. Analogous arguments apply to the study of acquisition of social information, where learning one behaviour can influence the likelihood of social learning of other behaviours. For example, extending models of information spreading (Aleta & Moreno, 2018; De Domenico et al., 2016; Kivelä et al., 2014) to two-aspect multilayer networks that include one layering aspect to represent different types of social interactions and another aspect to represent different time periods (Fig. 1) may provide valuable insights into how social dynamics influence cultural transmissions in a population.

The study of dispersal can also benefit from utilizing a multilayer framework. Networks have been used to uncover the role of spatial (Reichert, Fletcher, Cattau, & Kitchens, 2016) and social (Blumstein, Wey, & Tang, 2009) connectivity in dispersal decisions. One can use a two-aspect multilayer approach to integrate spatial layers that encode habitat connectivity or movements of individuals with social layers that encode intra-group and inter-group social relationships. For example, integrating a multilayer framework with existing multi-state models of dispersal (such as the ones in (Borg et al., 2017; Polansky, Kilian, & Wittemyer, 2015)) can make it possible to relate the likelihood of transitioning between dispersive and sedentary states to the positions of individuals in a multilayer socio-spatial network. Such integration of spatial and social contexts may provide new insights both into the relative roles of social and ecological environments in driving dispersal decisions and into the subsequent effects of dispersal on population structure.

### 2.3.1. Inter-specific interactions as a multilayer network

Network approaches have been useful for studying the social dynamics of mixed-species assemblages (Farine et al., 2012). For example, in mixed-species groups of passerine birds, network analysis was used to show that social learning occurs both within and between species (Farine, Aplin, Sheldon, & Hoppitt, 2015b). Mixed-species assemblages have an inherent multilayer structure. Most simply, one can represent a mixed-species community as a node-coloured network in which each layer represents a different species (Fig. 6). To incorporate additional useful information in a mixed-species multilayer network, one can represent the type of behavioural interaction as an additional aspect of the network. For example, one aspect can encode competitive interactions and another can encode non-competitive interactions.

Considering multilayer measures, such as multidegree or versatility, may provide new insights into the role of particular species or individuals in information sharing in mixed-species groups. Further, multilayer community detection has the potential to provide new insights into the structure of fission–fusion social systems that involve multiple species. The original study (Farine et al., 2015b) that generated the networks that we used in Fig. 6 investigated information transmission in both intra-species and inter-species social networks (i.e., constituent interaction types of an interconnected network). It concluded that both networks help predict the spread of information, but that the likelihood of acquiring foraging information was higher through intra-specific than through inter-specific associations, thereby providing a better understanding of information transmission in mixed-species communities than would be possible using monolayer network analysis. This highlights the potential of taking explicitly multilayer approaches to better understand how information can spread within and between species in mixed-species groups.

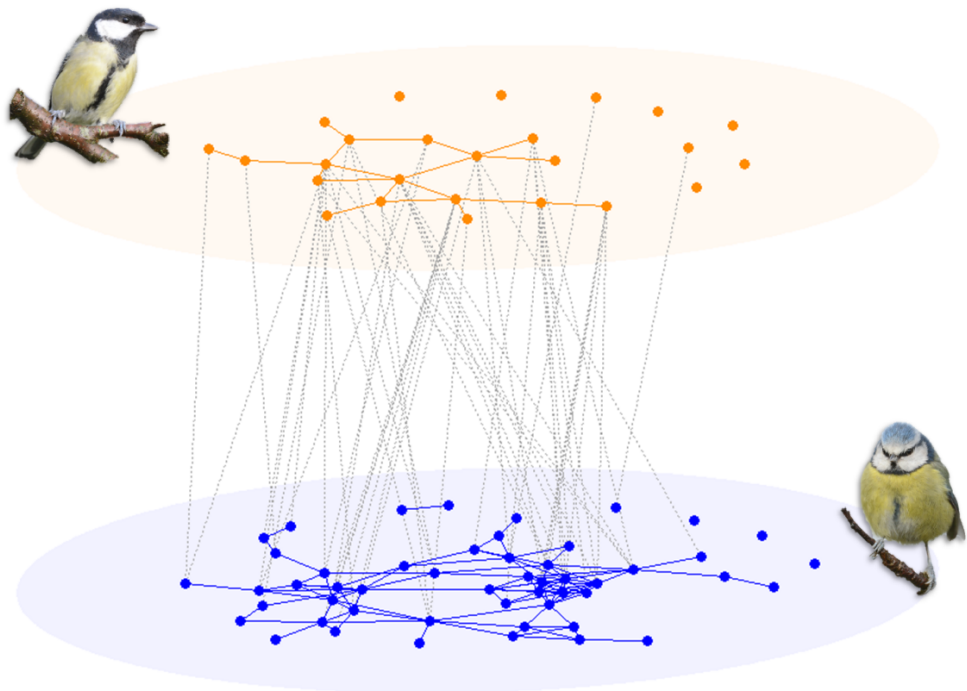

**Figure 6:** A multilayer network of mixed-species interactions between blue tits (bottom layer; blue nodes) and great tits (top layer; orange nodes) in Wytham Woods, UK (in the Cammoor–Stimpsons area) using data obtained from Dryad (Farine et al., 2015b; Farine, Aplin, Sheldon, & Hoppitt, 2015a). Each node represents an individual bird. Blue and orange edges connect individuals within layers (i.e., intra-specific associations), and grey edges connect individuals across layers (i.e., inter-specific associations). To aid clarity, we only show edges with a simple

ratio index (Cairns & Schwager, 1987; Ginsberg & Young, 1992) of 0.03 or larger. Photographs by Keith Silk.

**2.4 Evolutionary models**

Understanding the evolution of sociality is a central focus in evolutionary biology (Krause & Ruxton, 2002). Research approaches include agent-based simulations, game-theoretic models, comparative studies, and others. Evolutionary models have been expanded to incorporate interactions between agents, resulting in different evolutionary processes than those in models without interactions (Nowak, Tarnita, & Antal, 2010). However, social behaviours evolve and persist in conjunction with other behaviours and with ecological changes. Therefore, incorporating multiple types of interactions—social, physiological, and with an environment—as part of a multilayer framework can provide novel insights about the pressures on fitness and evolutionary processes. For example, incorporating interactions between molecules at the cellular level, organs at the organismal level, individuals at the group level, and groups at the population level into a network of networks can facilitate multilevel analysis of social evolution. In the ensuing paragraphs, we discuss how the expansion of evolutionary modelling approaches to include multilayer network analysis may enhance the study of (1) evolution of social phenomena (such as cooperation) and (2) co-variation in behavioural structures across species.

Incorporating ideas from network theory into evolutionary models has made it possible to account for long-term relationships, non-random interactions, and infrequent interactions (Lieberman, Hauert, & Nowak, 2005). These considerations can alter the outcomes of game-theoretic models of social evolution and facilitate the emergence or persistence of interactions, such as cooperation by enabling assortativity of cooperative individuals (Aktipis, 2004, 2006;

Allen et al., 2017; Croft, Edenbrow, & Darden, 2015; Fletcher & Doebeli, 2009; Nowak et al., 2010; Rand, Arbesman, & Christakis, 2011). Given the effects that group structure can have on the selection and stability of cooperative strategies, multilayer structures can significantly alter the dynamics (both outcomes and transient behaviour) of evolutionary games. Indeed, it has been demonstrated, using a multilayer network in which agents play games on multiple interconnected layers, that cooperation can persist under conditions in which it would not in a monolayer network (Gómez-Gardeñes, Reinares, Arenas, & Floría, 2012; Wang, Szolnoki, & Perc, 2012; Z. Wang, Wang, Szolnoki, & Perc, 2015). Furthermore, the level of interdependence, in the form of coupling payoffs between layers or by strategy transfer between layers, can influence the persistence of cooperation (Wang et al. 2013; Xia et al. 2014). Thus, in comparison to monolayer network analysis, using a multilayer network approach can improve the realism of models by better reflecting the 'multi-dimensional' nature of sociality and allowing a larger space of possible evolutionary strategies and outcomes. Certain behaviours that may not be evolutionarily stable when considering only one realm of social interactions may be able to evolve and/or persist when considering a multilayer structure of an agent's possible interactions. For example, expanding game-theoretic models to include multiple types of coupled interactions may facilitate the inclusion of both competition and mutualism, as well as both intra-specific and inter-specific interactions.

Comparative approaches offer another powerful method to examine the evolution of different social systems across similar species (Thierry, 2004; West-Eberhard, 1969). In socially complex species, such comparisons can benefit greatly from a multilayer approach. For instance, the macaque genus consists of over 20 species that exhibit a variety of different social structures, each with co-varying behavioural traits, such as those related to connectivity and/or individual behaviours (Thierry 2004; Sueur, Petit, et al. 2011; Balasubramaniam et al. 2012; Balasubramaniam et al. 2017). A multilayer network analysis of such co-varying interactions—e.g., with layers as connectivity types or time periods—may offer an effective way to reveal differences in social structure. For example, using matrix-correlation methods to measure similarities between layers in a multilayer network offers a way to compare how behaviours co-vary across different species using a multiple-regression quadratic assignment procedure (MRQAP) (Croft, Madden, Franks, & James, 2011). For multilayer networks, *global overlap* (Bianconi, 2013) and *global inter-clustering coefficient* (Parshani, Rozenblat, Ietri, Ducruet, & Havlin, 2010) are two measures that can quantify the overlap in edges between two layers. (See Appendix I for a brief discussion of layer-similarity measures.) One can, for instance, use global overlap between an affiliative network and a kinship network to examine the extent to which nepotism plays a role in social structure across species (Thierry, 2004). In such an analysis, it may also be useful to account for spatial dependencies.

Researchers continue to develop new approaches for measuring heterogeneous structures in multilayer networks (Aleta & Moreno, 2018; Kivelä et al., 2014) that can aid in testing specific evolutionary hypotheses. For example, the 'social-brain hypothesis' (Dunbar, 1998) posits that the evolution of cognition is driven by sociality, because sociality is cognitively challenging. Recently, there have been several propositions for how to define sociality to test the social-brain hypothesis; all of these include the idea that relationships between animals arise from different types of interactions (Bergman & Beehner, 2015; Fischer, Farnworth, Sennhenn-Reulen, & Hammerschmidt, 2017). Multilayer network analysis can aid in developing objective measures of social structures that include the nuances of the various proposed definitions. Another evolutionary hypothesis, the 'co-variation hypothesis' (Thierry, 2004), posits that changes in a single trait or behaviour can lead to changes in global social organization. Simulations of agent-based models (ABMs) on multilayer networks can test this hypothesis by exploring how different behavioural parameters along with coupling between layers influence group-level structure (Hemelrijk 2002). For example, an ABM of macaque societies (called 'Groofi world') linked grooming and fighting behaviour through a single trait (termed 'anxiety') (Hemelrijk & Puga-Gonzalez, 2012; Puga-Gonzalez, Hildenbrandt, & Hemelrijk, 2009). This model has an implicitly multilayer network structure, as it includes multiple interaction 'layers' that are coupled by a parameter. By incorporating such structure, the model illustrated that patterns of reciprocation and exchange (Hemelrijk & Puga-Gonzalez, 2012) and aggressive interventions (Puga-Gonzalez, Cooper, & Hemelrijk, 2016) can emerge from the presence of a few interconnected interaction types along with spatial positions.

## 3. Considerations when using multilayer network analysis

We have outlined many different opportunities for multilayer network approaches to be useful for the study of animal behaviour. However, the application of multilayer network analysis to animal behaviour data is in its infancy, with many exciting directions for future work. Multilayer network analysis may not always be appropriate for a given study, and there are several important considerations about both the applicability of the tools and the types of data on which to use them. Most importantly, practical implementation of these new tools will vary across study systems, and it will differ based on the questions asked. Therefore, researchers should not blindly implement these new techniques; instead, as with any other approach, they should be driven by their research questions and ensure that the tools and data are appropriate for answering those questions.

**3.1. When and how to use multilayer network analysis**

Multilayer network analysis adds complexity to the representation, analysis, and interpretation of data. Therefore, it should be applied only when incorporating a system's multifaceted nature can contribute to answering a research question, without adding needless complexity to data interpretation. Different types of social relationships may differ in the 'units' of their measurement, and it can be challenging to interpret multilayer network analysis of such integrated data. For example, if one layer represents genetic relatedness and another represents a social interaction, a multilayer similarity measure can reveal one or more relationships between these layers, but a versatility measure that uses both layers may be impractical or confusing to interpret, because they encode different types of connectivity data (i.e., relatedness and behaviour). In a similar vein, intralayer and interlayer edges can have entirely different meanings from each other, and it can sometimes be difficult to interpret the results of considering them jointly ((Kivelä et al., 2014); Appendix I).

Therefore, while the strength of using a multilayer network formalism is that it includes more information about interactions than a monolayer network, it is imperative to consider carefully which interactions to include in each layer, based on the study question. It is also important to be careful about which calculations are most appropriate for the different layers in a multilayer network, based on the functions of those layers, especially when they represent different behaviours.

**3.2. Data requirements**

Just as in monolayer network analysis (or in any study that samples a population), a key challenge is collecting sufficient and/or appropriately sampled data that provide a realistic depiction of the study system (Newman, 2018a, 2018b; Whitehead, 2008). Breaking data into multiple layers can result in sparse layers that do not provide an appropriate sample of the relationships in each layer. Further, if data sampling or sparsity varies across different layers or if the frequency of behaviours differs drastically, one layer may disproportionally dominate the outcome of a multilayer calculation. To avoid domination of one data type, one can threshold the associations, normalize edge weights, adjust interlayer edge weights (Appendix I), or aggregate layers (Appendix II) that include redundant information (De Domenico, Nicosia, Arenas, & Latora, 2015).

It is important to compare computations on a multilayer network to those on suitable randomizations (Farine, 2017; Kivelä et al., 2014). Just as in monolayer network analysis (Fosdick, Larremore, Nishimura, & Ugander, 2018; Newman, 2018c), it is important to tailor the use of null models in multilayer networks in a context-specific and question-specific way. For example, some network features may arise from external factors or hold for a large set of networks (e.g., all networks with the same intralayer degree distributions), rather than arising as distinctive attributes of a focal system.

### 3.3. Practical availability and further development of multilayer methodology

In practice, there are many ways for researchers in animal behaviour to implement multilayer network analysis. Existing software packages for examining multilayer networks include MuxViz (De Domenico, Porter, et al., 2015), Pymnet (Kivelä), and the R package Multinet (Magnani & Dubik, 2018). In Table 1, we summarize available tools for implementing various measures. Multilayer network analysis is a rapidly growing field of research in network science, and new measures and tools continue to emerge rapidly. Because this is a new, developing field of research, many monolayer network methods have not yet been generalized for multilayer networks; and many of the existing generalizations have not yet been implemented in publicly-available code. Additionally, many multilayer approaches have been published predominantly as proofs of concept in theoretically-oriented research or have been implemented only for multiplex networks, but not for other multilayer network structures (such as interconnected networks). Furthermore, multilayer networks with multiple aspects (e.g., time and behaviour type) have rarely been analysed in practice, and the potential utility of using multiple aspects to investigate questions about social behaviour may propel the development of tools to do so. The ongoing development of user-friendly software and modules is increasing the accessibility and practical usability of multilayer network analysis. Multilayer network analysis is very promising, but there is also a lot more work to do, as detailed above. Interdisciplinary collaborations between applied mathematicians, computer scientists, social scientists, behavioural ecologists, and others will be crucial for moving this exciting new field forward.

## 4. Conclusions

In this article, we have discussed the use of multilayer network analysis and outlined potential uses for providing insights into social behaviour in animals. Multilayer networks provide a useful framework for considering many extensions of animal social network analysis. For example, they make it possible to incorporate temporal and spatial processes alongside multiple types of behavioural interactions in an integrated way. We have highlighted examples in which multilayer methods have been used previously to study animal behaviour, illustrated them with several case studies, proposed ideas for future work in this area, and provided practical guidance on some suitable available methodologies and software (Table 1). Using multilayer network analysis offers significant potential for uncovering eco-evolutionary dynamics of animal social behaviour. Multilayer approaches provide new tools to advance research on the evolution of sociality, group and population dynamics, and the roles of individuals in interconnected social and ecological systems. The incorporation of multilayer methods into studies of animal behaviour will facilitate an improved understanding of what links social dynamics across behaviours and contexts, and it also provides an explicit framework to link social behaviour with broader ecological and evolutionary processes (Silk, Finn, Porter & Pinter-Wollman, 2018).

**Table 1:** A non-exhaustive selection of multilayer network approaches for studying questions in behavioural ecology. We provide a description of each tool and point to software in which they are implemented. We note the organizational level(s) (individual (I), group (G), population (P), and evolution (E)) of the tools. We provide examples of questions that can be investigated with each approach. These questions provide general guidelines for more specific hypotheses that would be guided by the study system and biological questions of interest.

| Research aim | Level (I/G/P/E) | Examples of questions | Multilayer approach | Description | Software package | Citation |
|---|---|---|---|---|---|---|
| Identify important or influential nodes or edges | I/G | • How will a group be affected if certain individuals are removed?<br>• Is social influence determined by interactions in more than one situation?<br>• Which relationships are most critical for group cohesion (when applying measures to edges)?<br>• How stable is an individual's importance over time? | Eigenvector versatility | Multilayer extension of eigenvector centrality, for which an individual's importance depends on its connections within and across layers and on the connections of its neighbours. | MuxViz (De Domenico, Porter, et al., 2015) | (De Domenico, Solé-Ribalta, et al., 2015) |
| | | • Which individuals link the most individuals in a group within or across social situations and/or over time?<br>• How important is an individual for group cohesion? | Betweenness versatility | Multilayer extension of geodesic betweenness centrality, which measures how often shortest paths (including both intralayer and interlayer edges) between each pair of nodes traverse a given node. | MuxViz | (De Domenico, Solé-Ribalta, et al., 2015) |
| | | • Does the role of an individual in its social group carry over across social situations? | Multidegree | A vector of the intralayer degrees of each individual across all layers | Pymnet (Kivelä, n.d.) | (Menichetti, Remondini, Panzarasa, Mondragón, & Bianconi, 2014) |
| Quantify network properties at different scales | G/P/E | | • What are the coherent groups in a network of animals?<br>• Which individuals preferentially interact with each other in different or multiple contexts? | Multislice modularity maximization, Multilayer InfoMap | Identifies communities of individuals in which the same individuals in different layers can be assigned to different communities. | MuxViz; GenLouvain: https://github.com/GenLouvain/GenLouvain | (Mucha et al., 2010) |
| | | | • What are the social communities, core–periphery structures, or other large-scale structures in different types of social situations? | Stochastic block models | Statistical models of arbitrary block structures in networks. | Graph-tool (Python) | (Peixoto, 2015) |
| | | | • Are there consistent, 'typical' types of interaction patterns across social situations? | Motifs | Interaction patterns between multiple individuals (e.g., node pairs or triples), within and/or across layers, that appear more often than in some null model. | MuxViz | (Battiston, Nicosia, Chavez, & Latora, 2017; Wernicke & Rasche, 2006) |
| | | | • How similar are the interaction patterns in different social situations?<br>• How often do interactions between individuals co-occur in multiple situations? | Global overlap | Number of pairs of nodes that are connected by edges in multiple layers. | MuxViz; Multinet R package (Magnani & Dubik, 2018) | (Bianconi, 2013) |
| Model statistical properties of a network | G/P/E | | • Are interaction patterns influenced by group size? | Randomization for multilayer networks | Construction of randomized ensembles of synthetic multilayer networks for comparison | Pymnet (Python) | (Kivelä et al., 2014), Section 4.3 |
| | | | • Are relationships or interactions in one social situation related to relationships or interactions in a different social situation? | Exponential random graph model (ERGM) | An extension of ERGMs to multilayer networks | MPNET (Java-based) for | (Heaney, 2014; P. Wang, |

| | | | | | | |
|---|---|---|---|---|---|---|
| | | • Are relationships at one time point related to those at a different time point? | | | two-layer multilayer networks | Robins, Pattison, & Lazega, 2013) |
| | | • How do network relationships in one social situation or at one point in time affect subsequent relationships in other situations or at other times? | Markov models of co-evolving multiplex networks | Models of the probability of an edge existing in a layer at one time as a function of an edge existing between the same pair of nodes in any layer in the previous time. | Multiplex MarkovChain: https://github.com/vkrmsv/MultiplexMarkovChain | (Fisher et al., 2017; Vijayaraghavan, Noël, Maoz, & D'Souza, 2015) |
| | | | Stochastic actor-oriented models for multiple networks | Statistical models of what influences the creation and termination of edges between times. The version that we consider can model the co-evolution of two networks (or two layers) as a result of their influence on each other. | Code available at https://www.stats.ox.ac.uk/~snijders/siena/siena_scripts.htm | |

| Modeling disease or information transmission | I/G/P | <ul><li>What are the roles of different types of social interactions or individuals in information or disease transmission?</li><li>Do different types of transmission interact or interfere with each other?<ul><li>For example, can the spread of information mitigate the spread of a disease?</li><li>Can the spread of one infection enhance or reduce the spread of a second infection?</li></ul></li><li>What influences disease transmission in multi-species communities?</li></ul> | Compartmental models on networks | Classic epidemiological models that assume that individuals exist in one of several states, with probabilistic transitions between states. For example, SIR models have susceptible, infective, and recovered (or removed) states; and SI and SIS models have only susceptible and infected states. These models are sometimes amenable to mathematical analysis, but stochastic simulations are often more accessible. | EpiModel (R package) (temporal multiplex networks only) ("EpiModel," n.d.; Jenness, Goodreau, & Morris, 2017) | (Pastor-Satorras et al. 2015; Kiss et al. 2017; Porter and Gleeson 2016) |

**Acknowledgements**

We thank MX16 (Multidimensional Networks Symposium 2016, University of California, Davis) co-organizers Curtis Atkisson and Jordan Snyder, as well as the MX16 participants, for inspiring thoughts and conversations about multilayer networks that helped instigate this collaboration. We also thank Raissa D'Souza and the members of her lab for discussions on multilayer networks and Brenda McCowan and her lab members, especially Brianne Beisner, for support and extensive conversations about the 'multi-dimensionality' of macaque societies. We thank Haochen Wu for assisting with installation of software modules in exchange for beer. Finally, we thank Tiago de Paula Peixoto and Manlio De Domenico for helpful discussions on the statistical modelling of multilayer networks.

**Funding**

KRF was funded by the National Science Foundation (NSF) Graduate Research Fellowship (1650042). NPW was funded by NFS IOS grant 1456010/1708455 and NIH R01 GM115509. MJS was funded by NERC standard grant NE/M004546/1. We gratefully acknowledge the supporters of MX16: the UC Davis Institute for Social Sciences, the U.S. Army Research Office under Multidisciplinary University Research Initiative Award No. W911NF-13-1-0340, the UC Davis Complexity Sciences Center, the UC Davis Anthropology Department, the UC Davis Graduate Student Association, the UC Davis Department of Engineering, and the UC Davis Office of Research.

Carter, D. R., DeChurch, L. A., Braun, M. T., & Contractor, N. S. (2015). Social network approaches to leadership: An integrative conceptual review. *Journal of Applied Psychology*, *100*(3), 597–622. https://doi.org/10.1037/a0038922

Chan, S., Fushing, H., Beisner, B. A., & McCowan, B. (2013). Joint modeling of multiple social networks to elucidate primate social dynamics: I. Maximum entropy principle and network-based interactions. *PLoS ONE*, *8*(2), e51903. https://doi.org/10.1371/journal.pone.0051903

Chodrow, P. S., Al-Awwad, Z., Jiang, S., González, M. C., Herranz, R., & Frias-Martinez, E. (2016). Demand and congestion in multiplex transportation networks. *PLOS ONE*, *11*(9), e0161738. https://doi.org/10.1371/journal.pone.0161738

Couzin, I. D., Krause, J., James, R., Ruxton, G. D., & Franks, N. R. (2002). Collective Memory and Spatial Sorting in Animal Groups. *Journal of Theoretical Biology*, *218*(1), 1–11. https://doi.org/10.1006/jtbi.2002.3065

Couzin, I. D., & Laidre, M. E. (2009). Fission–fusion populations. *Current Biology*, *19*(15), R633–R635. https://doi.org/10.1016/j.cub.2009.05.034

Craft, M. E. (2015). Infectious disease transmission and contact networks in wildlife and livestock. *Philosophical Transactions of the Royal Society B: Biological Sciences*, *370*(1669), 20140107–20140107. https://doi.org/10.1098/rstb.2014.0107

Croft, D. P., Edenbrow, M., & Darden, S. K. (2015). Assortment in social networks and the evolution of cooperation. In *Animal Social Networks* (pp. 13–23).

Croft, D. P., Edenbrow, M., Darden, S. K., Ramnarine, I. W., van Oosterhout, C., & Cable, J. (2011). Effect of gyrodactylid ectoparasites on host behaviour and social network structure in guppies Poecilia reticulata. *Behavioral Ecology and Sociobiology*, *65*(12), 2219–2227. https://doi.org/10.1007/s00265-011-1230-2
50

96. https://doi.org/10.1016/j.anbehav.2012.10.010

Ilany, A., Booms, A. S., & Holekamp, K. E. (2015). Topological effects of network structure on long-term social network dynamics in a wild mammal. *Ecology Letters*, *18*(7), 687–695. https://doi.org/10.1111/ele.12447

Jack, K. M. (2003). Explaining variation in affiliative relationships among male White-Faced Capuchins (Cebus capucinus). *Folia Primatol*, *74*, 1–16. https://doi.org/10.1159/000068390

Jenness, S., Goodreau, S. M., & Morris, M. (2017). EpiModel: An R package for mathematical modeling of infectious disease over etworks. *BioRxiv*, 213009. https://doi.org/10.1101/213009

Jeub, L. G. S., Balachandran, P., Porter, M. A., Mucha, P. J., & Mahoney, M. W. (2015). Think locally, act locally: Detection of small, medium-sized, and large communities in large networks. *Physical Review E*, *91*(1), 012821. https://doi.org/10.1103/PhysRevE.91.012821

Jeub, L. G. S., Mahoney, M. W., Mucha, P. J., & Porter, M. A. (2017). A local perspective on community structure in multilayer networks. *Network Science*, *5*(2), 144–163. https://doi.org/10.1017/nws.2016.22

Kasper, C., & Voelkl, B. (2009). A social network analysis of primate groups. *Primates*, *50*(4), 343–356. https://doi.org/10.1007/s10329-009-0153-2

Kiss, I. Z., Miller, J. C., & Simon, P. L. (2017). *Mathematics of Epidemics on Networks* (Vol. 46). Cham: Springer International Publishing. https://doi.org/10.1007/978-3-319-50806-1

Kivelä, M. (n.d.). Pymnet: Mulilayer networks library for Python. Retrieved from http://www.mkivela.com/pymnet/

Kivelä, M., Arenas, A., Barthélemy, M., Gleeson, J. P., Moreno, Y., & Porter, M. A. (2014). Multilayer networks. *Journal of Complex Networks*, *2*(3), 203–271.
56

Grevy's Zebra. *PLOS ONE*, *10*(10), e0138645.

https://doi.org/10.1371/journal.pone.0138645

Shizuka, D., Chaine, A. S., Anderson, J., Johnson, O., Laursen, I. M., & Lyon, B. E. (2014). Across-year social stability shapes network structure in wintering migrant sparrows. *Ecology Letters*, *17*(8), 998–1007. https://doi.org/10.1111/ele.12304

Sih, A., Bell, A., & Johnson, J. C. (2004). Behavioral syndromes: An ecological and evolutionary overview. *Trends in Ecology and Evolution*, *19*(7), 372–378. https://doi.org/10.1016/j.tree.2004.04.009

Sih, A., Hanser, S. F., & McHugh, K. A. (2009). Social network theory: New insights and issues for behavioral ecologists. *Behavioral Ecology and Sociobiology*, *63*(7), 975–988. https://doi.org/10.1007/s00265-009-0725-6

Silk, J. B., Alberts, S. C., & Altmann, J. (2003). Social bonds of female baboons enhance infant survival. *Science*, *302*(5648), 1231–1234. https://doi.org/10.1126/science.1088580

Silk, M. J., Croft, D. P., Tregenza, T., & Bearhop, S. (2014). The importance of fission-fusion social group dynamics in birds. *Ibis*, *156*(4), 701–715. https://doi.org/10.1111/ibi.12191

Silk, M. J., Finn, K. R., Porter, M. A., & Pinter-Wollman, N. (2018). Forum: Can multilayer networks advance animal behavior research? *Trends in Ecology & Evolution*, *33*(6), 376–378.

Silk, M. J., Weber, N. L., Steward, L. C., Hodgson, D. J., Boots, M., Croft, D. P., … McDonald, R. A. (2018). Contact networks structured by sex underpin sex-specific epidemiology of infection. *Ecology Letters*, *21*(2), 309–318. https://doi.org/10.1111/ele.12898

Smith-Aguilar, S. E., Aureli, F., Busia, L., Schaffner, C., & Ramos-Fernández, G. (2018). Using multiplex networks to capture the multidimensional nature of social structure. *Primates*, 1–
62

# APPENDIX I OF 'THE USE OF MULTILAYER NETWORK ANALYSIS ACROSS SOCIAL SCALES IN ANIMAL BEHAVIOUR'

KELLY R. FINN, MATTHEW J. SILK, MASON A. PORTER, AND NOA PINTER-WOLLMAN

The mathematical formalism of *multilayer networks* [Kivelä et al., 2014, Newman, 2018], a generalization of ordinary graphs (i.e., 'monolayer networks'), was developed recently to help study multitudinous types of networks and to unify them into one framework. In this appendix, we complement the main text with an introduction to this formalism. We follow the mathematical approach of the review article [Kivelä et al., 2014], including most of their terminology and much of their notation. Another useful resource is [Porter, 2018], which is an expository summary of multilayer networks for mathematics students.[1]

## 1. Mathematical Formalism

A multilayer network $M = (V_M, E_M, V, \texttt{L})$ has an underlying set $V = \{1, \ldots, N\}$ of $N$ entities (i.e., 'physical nodes') that occur on layers in $\texttt{L}$, which we construct as a sequence, $\texttt{L} = \{L_a\}_{a=1}^d$, of sets $L_1, \ldots, L_d$ of elementary layers, where $d$ is the number of 'aspects' (i.e., types of layering). One 'layer' in $\texttt{L}$ is thus a combination, through the Cartesian product $L_1 \times \cdots \times L_d$, of 'elementary layers' from all aspects. Therefore, each layer in a multilayer network includes one elementary layer from each aspect. For example, if the sets of elementary layers of a multilayer network are $L_1 = \{1, 2\}$ (so 1 and 2 are each elementary layers, perhaps representing different points in time) and $L_2 = \{X, Y, Z\}$, then the network's layers are $(1, X)$, $(1, Y)$, $(1, Z)$, $(2, X)$, $(2, Y)$, and $(2, Z)$. In a multilayer network, the set of node-layer tuples (i.e., 'state nodes' that correspond to the same entity) in $M$ is $V_M \subseteq V \times L_1 \times \cdots \times L_d$, and the set of multilayer edges is $E_M \subseteq V_M \times V_M$. The edge $((i, \alpha), (j, \beta)) \in E_M$ indicates that there is an edge from node $i$ on layer $\alpha$ to node $j$ on layer $\beta$ (and vice versa, if $M$ is undirected). Each aspect of $M$ represents a type of layering: a type of social interaction, a point in time, and so on. For example, a multirelational network that does not change in time has one aspect; a multirelational network that has layers that encompass multiple time points has two aspects; and so on. To consider weighted edges, one proceeds as in monolayer networks by assigning a weight to each edge using a function $w : E_M \longrightarrow X$, where $X = \mathbb{R}_{\geq 0}$ if all weights are nonnegative real numbers. In Fig. S1, we show an example of a multilayer network with one aspect.

1.1. **Adjacency Structure.** Each multilayer network with the same number of nodes in each layer has an associated adjacency tensor[2] $\mathcal{A}$ of order $2(d+1)$. See [Kivelä et al., 2014] for details. Analogous to the case of monolayer networks, each unweighted (and directed[3]) edge in $E_M$ is associated with a 1 entry of $\mathcal{A}$, and the other entries are 0. To incorporate edge weights, one uses the values of the weights instead of 1. As discussed in [Kivelä et al., 2014], multilayer networks can have different numbers of nodes in different layers. To ensure that the dimensions are consistent in $\mathcal{A}$, one adds empty state nodes when necessary. Edges attached to such state nodes are 'forbidden' (these yield 'structural zeros' in $\mathcal{A}$), and this needs to be taken into account when doing calculations.

---

[1]We draw on some exposition from [Porter, 2018] in our section on mathematical formalism.
[2]A tensor is a linear-algebraic object that generalizes a matrix. See [De Domenico et al., 2013, Kivelä et al., 2014] for discussions of tensors in the context of multilayer networks.
[3]An undirected edge in $E_M$ is associated with two 1 entries of $\mathcal{A}$.

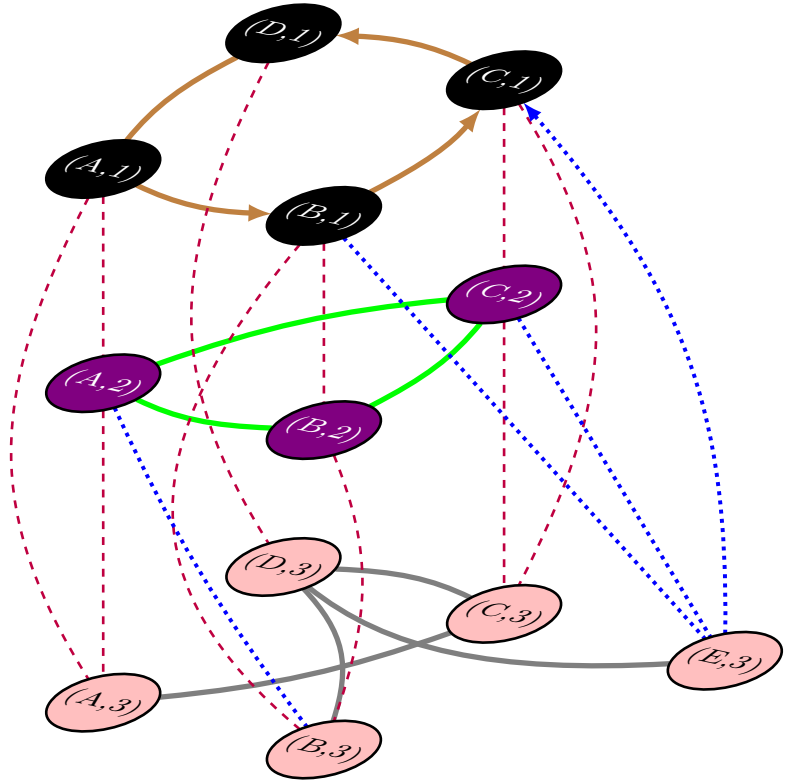

FIGURE S1. An example of a multilayer network with three layers. We label each layer using different colours for its state nodes and its edges: black nodes and brown edges (three of which are unidirectional) for layer 1, purple nodes and green edges for layer 2, and pink nodes and grey edges for layer 3. Each state node (i.e., node-layer tuple) has a corresponding physical node and layer, so the tuple $(A, 3)$ denotes physical node $A$ on layer 3, the tuple $(D, 1)$ denotes physical node $D$ on layer 1, and so on. We draw intralayer edges using solid arcs and interlayer edges using broken arcs; an interlayer edge is dashed (and magenta) if it connects corresponding entities and dotted (and blue) if it connects distinct ones. We include arrowheads to represent unidirectional edges. [We drew this network using TIKZ-NETWORK, by Jürgen Hackl and available at https://github.com/hackl/tikz-network), which allows one to draw multilayer networks directly in a LaTeX file.]

For convenience[4], it is common to flatten $\mathcal{A}$ into a 'supra-adjacency matrix' $\mathbf{A}_M$, which is the adjacency matrix of the graph $G_M$ associated with $M$. Intralayer edges (the solid arcs in Fig. S1) are on the diagonal blocks of a supra-adjacency matrix

---

[4]Several developments in multilayer network analysis, including particular choices for how to generalize ideas from monolayer network analysis, have exploited the tensorial structure of multilayer networks. Readers who wish to ignore this structure are free to start with supra-adjacency matrices.

$$\mathbf{A}_M = \begin{pmatrix} \begin{array}{ccccc|ccccc|ccccc} 0 & 1 & 0 & 1 & 0 & \omega_{A1,A2} & 0 & 0 & 0 & 0 & \omega_{A1,A3} & 0 & 0 & 0 & 0 \\ 0 & 0 & 1 & 0 & 0 & 0 & \omega_{B1,B2} & 0 & 0 & 0 & 0 & \omega_{B1,B3} & 0 & 0 & \omega_{B1,E3} \\ 0 & 0 & 0 & 1 & 0 & 0 & 0 & \omega_{C1,C2} & 0 & 0 & 0 & 0 & \omega_{C1,C3} & 0 & 0 \\ 1 & 0 & 0 & 0 & 0 & 0 & 0 & 0 & 0 & 0 & 0 & 0 & 0 & \omega_{D1,D3} & 0 \\ 0 & 0 & 0 & 0 & 0 & 0 & 0 & 0 & 0 & 0 & 0 & 0 & 0 & 0 & 0 \\ \hline \omega_{A2,A1} & 0 & 0 & 0 & 0 & 0 & 1 & 1 & 0 & 0 & \omega_{A2,A3} & \omega_{A2,B3} & 0 & 0 & 0 \\ 0 & \omega_{B2,B1} & 0 & 0 & 0 & 1 & 0 & 1 & 0 & 0 & 0 & \omega_{B2,B3} & 0 & 0 & 0 \\ 0 & 0 & \omega_{C2,C1} & 0 & 0 & 1 & 1 & 0 & 0 & 0 & 0 & 0 & \omega_{C2,C3} & 0 & \omega_{C2,E3} \\ 0 & 0 & 0 & 0 & 0 & 0 & 0 & 0 & 0 & 0 & 0 & 0 & 0 & 0 & 0 \\ 0 & 0 & 0 & 0 & 0 & 0 & 0 & 0 & 0 & 0 & 0 & 0 & 0 & 0 & 0 \\ \hline \omega_{A3,A1} & 0 & 0 & 0 & 0 & \omega_{A3,A2} & 0 & 0 & 0 & 0 & 0 & 0 & 1 & 0 & 0 \\ 0 & \omega_{B3,B1} & 0 & 0 & 0 & \omega_{B3,A2} & \omega_{B3,B2} & 0 & 0 & 0 & 0 & 0 & 0 & 1 & 0 \\ 0 & 0 & \omega_{C3,C1} & 0 & 0 & 0 & 0 & \omega_{C3,C2} & 0 & 0 & 1 & 0 & 0 & 1 & 0 \\ 0 & 0 & 0 & \omega_{D3,D1} & 0 & 0 & 0 & 0 & 0 & 0 & 0 & 1 & 1 & 0 & 1 \\ 0 & \omega_{E3,B1} & \omega_{E3,C1} & 0 & 0 & 0 & 0 & \omega_{E3,C2} & 0 & 0 & 0 & 0 & 0 & 1 & 0 \end{array} \end{pmatrix}$$

FIGURE S2. Supra-adjacency matrix corresponding to the multilayer network in Fig. S1. Entries in diagonal blocks correspond to intralayer edges, whereas entries in off-diagonal blocks correspond to interlayer edges. We follow the colouring scheme in Fig. S1: entries that correspond to intralayer edges in layer 1 are in brown, those in layer 2 are in green, and those in layer 3 are in grey. Magenta entries correspond to interlayer edges between state nodes that represent the same entity, and blue entries correspond to interlayer edges between state nodes that represent distinct entities. We use subscripts to identify the weights of the specific interlayer edges; for example, $\omega_{A1,A2}$ denotes the weight of the edge from state node $(A, 1)$ to state node $(A, 2)$, and $\omega_{A2,A1}$ denotes the weight of the edge from $(A, 2)$ to $(A, 1)$. As in monolayer networks, intralayer edges can also be weighted, but we do not indicate any such weights in $\mathbf{A}_M$.

(see Fig. S2), and interlayer edges (the dashed magenta arcs and dotted blue arcs in Fig. S1) are on the off-diagonal blocks. For this illustration, we suppose that the intralayer edges are unweighted; these are the 1 entries (which we colour based on layer) in $\mathbf{A}_M$. We show interlayer edges between state nodes that represent the same entity in magenta, and we show interlayer edges between state nodes that represent distinct entities in blue. We suppose that the interlayer edge from state node $(i, \alpha)$ to $(j, \beta)$ has weight $\omega_{i\alpha,j\beta}$, which we take to be a nonnegative real number (although one can use negative values for antagonistic interlayer interactions).

1.2. **Types of Multilayer Networks.** Multilayer networks allow one to investigate a diverse variety of complicated network architectures and to integrate different types of data into one mathematical object. One can then use a common toolkit to study these diverse scenarios.

Two key types of multilayer networks arise from (i) labeling edges or (ii) labeling nodes. When one labels edges, one thinks of edges in different layers as representing different types of interactions. This is the case for a *multiplex network*, a type of multilayer network in which the only permitted types of interlayer edges are those that connect replicates of the same entity in different layers. We show such edges, which correspond to diagonal elements of off-diagonal blocks in a supra-adjacency matrix, as dashed magenta arcs in Fig. S1 and as magenta matrix elements in Fig. S2. A special case of a multiplex network is an edge-coloured multigraph, which has multiple layers but does not have any interlayer edges. In this case, only the diagonal blocks in a supra-adjacency matrix can have nonzero elements (such that all $\omega_{i\alpha,j\beta} = 0$ in Fig. S2). By contrast, when one labels nodes, one can think of different layers as representing different subsystems (in *interconnected networks* and 'networks of networks'), and there can be interlayer edges with nonzero supra-adjacency matrix elements in both the diagonal and off-diagonal entries of the off-diagonal blocks. In this case, there are interlayer edges between different entities, as we indicate using the dotted blue arcs in Fig. S1 and the blue matrix elements in Fig. S2. To emphasize the fact that different layers in Fig. S1 can represent different subsystems, we use different colours for the nodes from different layers in this network diagram.

For further details on types of multilayer networks, see [Kivelä et al., 2014].

1.3. **Weights of Interlayer Edges.** An important idea is that interlayer edges are fundamentally different from intralayer edges, and it is often less straightforward to determine weights from data for interlayer edges than for intralayer ones. In the context of the supra-adjacency matrix in Fig. S2, for most applications, it is easier to determine weights that are associated with the 1 entries in the diagonal blocks than to assign appropriate values to the weights $\omega_{i\alpha,j\beta}$. As in monolayer networks, larger weights correspond to stronger interactions.

A conceptually easy situation is a multimodal transportation network, in which one might determine interlayer edge weights based on how long it takes to change modes of transportation (with larger weights for shorter times). Suppose, for example, that entity $A$ represents Oxford, entity $B$ represents Cambridge, layer 1 represents coach transportation, and layer 2 represents train transportation. We determine interlayer edge weights from the time it takes to change transportation modes, with larger weights for shorter times. If it takes longer to walk from the coach station to the train station in Oxford than it does in Cambridge, then $\omega_{A1,A2} > \omega_{B1,B2}$.

A harder scenario to model is communication between people in a social network. We will use ourselves—with nodes called Mason, Noa, Kelly, and Matt—to provide an example. One possibility is to construe an interlayer edge that connects an entity to itself as encoding a transition probability between different modes of communication. Therefore, $\omega_{i\alpha,j\beta} \in [0,1]$ because it represents a probability. One can also include interlayer edges between distinct entities (in blue in Fig. S2), as Mason can send a message to Noa using one mode of communication (i.e., in one layer), such as via an e-mail that he typed on his laptop, but she may read the contents of that message using some other mode of communication (i.e., in another layer), such as on a mobile phone. Noa may then subsequently text the message to Kelly and Matt. Additionally, because the four of us have different usage patterns for different modes of communication, we also have different transition probabilities between layers, and our associated interlayer edge weights thus differ from each other. For example, Mason is almost always on his computer and almost never on his phone, so his transition probability from communicating via computer to communicating via phone is small, whereas the probability of the reverse transition is very large. By contrast, during proverbial work hours, Noa spends a similar amount of time on her computer and her phone, and her transition probabilities for changing between these two modes of communication are similar to each other.

For other applications, including in animal behaviour, interlayer edges can run into significant conceptual difficulties, and researchers struggle with how to make sense of them. There are dependencies across layers and interlayer edges can encode such dependencies, but how does one determine meaningful values for the weights of those edges? In some applications, it may be useful to think of interlayer edges as transition probabilities, as in the above example involving humans. In others, it may be useful to construe an interlayer edge as representing a dependency between one layer (e.g., proximity associations) and a second layer (e.g., grooming interactions, which require proximity to occur). A larger weight for such an edge encodes a stronger dependency, thereby entailing a stronger dependence of one layer on another. Additionally, different individual animals can have different values for such weights (as in the example above), corresponding to individual differences.

There are numerous possibilities for applying multilayer network analysis in animal behaviour (and in other applications), because it is very flexible, but it can also be very challenging to interpret the results of such analysis. As we have illustrated in this subsection, a key issue that requires careful thought is determining the weights of interlayer edges (or whether to use such edges at all). In different disciplines and for different systems and research questions, one can use interlayer edges to represent qualitatively different things (e.g., communication ties, correlations, or transition probabilities), and how to determine interlayer edge weights depends on the application domain, the system of interest, and one's particular research question.

## 2. Eigenvector Versatility: An Example of a Multilayer Versatility Measure

In this section, we illustrate the formalism of calculating a 'versatility' measure [De Domenico et al., 2015] in a multilayer network to supplement our conceptual discussion in the main text. For simplicity, we consider eigenvector versatility, which is a generalization of eigenvector centrality from monolayer networks, but one can also generalize other monolayer centrality measures (such as PageRank) into associated versatility measures for multilayer networks.

To calculate eigenvector centralities in a monolayer network, one calculates the leading eigenvector $\mathbf{v}_1$ (which is associated with the largest positive eigenvalue $\lambda_1$) of the equation $\mathbf{A}\mathbf{v} = \lambda\mathbf{v}$, where $\mathbf{A}$ is the network's adjacency matrix. For this type of centrality, we assume that the network associated to $\mathbf{A}$ is strongly connected (or just that it is connected, for an undirected network), so that—by the Perron–Frobenius theorem—the eigenvector $\mathbf{v}_1$ has strictly positive entries [Newman, 2018]. These entries give the eigenvector centralities of the corresponding nodes in the network.

Calculating eigenvector versatility proceeds in a similar way. One first calculates the leading eigenvector $\mathbf{v}_{M,1}$ of the equation $\mathbf{A}_M\mathbf{v}_M = \lambda\mathbf{v}_M$. The eigenvector $\mathbf{v}_{M,1}$ gives multilayer eigenvector centralities for each state node (i.e., for each node in each layer). Importantly, we need to use the whole multilayer structure to calculate the multilayer eigenvector centrality for each state node. For each entity, one then aggregates the centrality values over all layers to determine its eigenvector versatility. The article [De Domenico et al., 2015] used a maximum-entropy principle for their choice of aggregation, but other ways of weighting different layers are also possible [Kivelä et al., 2014].

## 3. Similarity of Layers: An Example Measure

In this section, we present one example measurement of similarity of layers in a multilayer network. As we suggested in the main text, such calculations can be helpful for exploring overlaps of individuals and/or social interactions across layers, including for discerning task specialists and generalists. For simplicity, we consider the special case of multiplex networks. See [Kao and Porter, 2018] and several references therein for a discussion of several types of similarity measures and comparisons between them.

One way to quantify the similarity of two layers is to count the number of intralayer edges that occur in both layers. There is an overlapping edge between nodes $i$ and $j$ in layers $\alpha$ and $\beta$ if and only if there is an edge between nodes $i$ and $j$ in both $\alpha$ and $\beta$. That is, $\theta(A_{ij}^\alpha) = 1$ and $\theta(A_{ij}^\beta) = 1$, where $A_{ij}^\alpha$ is the intralayer adjacency element between entities $i$ and $j$ on layer $\alpha$ (and $A_{ij}^\beta$ is defined analogously), and $\theta(x) = 1$ if $x > 0$ and $\theta(x) = 0$ otherwise.

Usually, one wants to be a bit more sophisticated than using a raw count of overlapping edges, and there are many possible ways to proceed. One example is 'local overlap' [Cellai et al., 2013]

$$o_i^{\alpha\beta} = \sum_j \theta(w_{ij}^\alpha)\theta(w_{ij}^\beta)\,,$$

which counts the number of overlapping edges that are attached to node $i$ in both layer $\alpha$ and layer $\beta$. In an undirected multiplex network, the local overlap $o_i^{\alpha\beta}$ quantifies the similarity between the connection patterns of node $i$ in layer $\alpha$ and node $i$ in layer $\beta$, although it does not take into account that the intralayer degrees of a state node contributes to the amount of overlap that involves it. One way to do this is with 'local similarity' [Kao and Porter, 2018]

$$\phi_i^{\alpha\beta} = \frac{o_i^{\alpha\beta}}{k_i^\alpha + k_i^\beta - o_i^{\alpha\beta}} \in [0,1]\,, \tag{1}$$

where $k_i^\alpha = \sum_j \theta(w_{ij}^\alpha)$ is the degree of node $i$ in layer $\alpha$ (and $k_i^\beta$ is defined analogously). Local similarity $\phi_i^{\alpha\beta}$ calculates the number of overlapping edges that are attached to node $i$ in layers $\alpha$ and $\beta$ as a proportion of the number of unique edges that are attached to node $i$ in the two layers.

# APPENDIX II

## 1. Representation of multifaceted data as a monolayer network

One approach that has been utilized widely for studying multifaceted systems is to aggregate 'layers' of different data types into a weighted monolayer network and to then use traditional network measures to quantify properties of the resulting aggregated network. The simplest aggregation is a linear combination in which each edge between two individuals in a layer adds (or subtracts) from the weight of the edge in the aggregated monolayer network. For example, if we assign grooming interactions (affiliative) a value of +1 and aggressive interactions (agonistic) a value of -1 and combine them through simple addition, then two individuals with 6 agonistic interactions and 1 grooming interaction have an edge of weight -5 in the aggregated network. Another approach is to aggregate in a way that assigns a greater importance to a particular layer. In the above example, if we decide (for example, by using pre-existing knowledge of a study system) that grooming interactions are twice as important as agonistic interactions in defining social relationships, the monolayer edge between the above individuals in the ensuing aggregation now gets a weight of -4. Therefore, aggregating multifaceted data that has negatively weighted edges into a monolayer network yields a weighted signed network, for which there are fewer methods of analysis than there are for networks in which edge weights are nonnegative (Traag, Doreian, & Mrvar, 2018; Wasserman & Faust, 1994). It is possible to adjust edge weights so that they are all positive, although one has to be careful about how this changes the information that is encoded in a network. Additionally, the relative importance of each layer is often not known, and it is not necessarily appropriate to weight them in linear proportion.

## 2. Tuning interlayer coupling with relaxation rates

When detecting communities in a network using the multilayer InfoMap algorithm, one can tune the amount of coupling between layers (i.e., the influence of interlayer edges on the communities that are detected) by adjusting the 'relaxation rate' of a random walk. This relaxation rate determines how much interlayer edges (with uniform values of 1, using default settings of MuxViz) are taken into account when detecting communities; a value of 0 entails no influence of interlayer edges on community assignment (no coupling), and a value of 1 entails that there is no distinction between layers (i.e., the assumption of distinction is 'relaxed') (De Domenico, Lancichinetti, Arenas, & Rosvall, 2015). Depending on the study system, the specific function of intralayer and interlayer edges, and the study-specific research aim for examining communities, one may wish to use different relaxation rates or to explore how community composition changes across a range of relaxation rates (Fig. S3).

For instance, it may be appropriate to use a small relaxation rate for a multiplex network in which the interactions in different layers are functionally very different (e.g., associating at work, at a bar, or at home) and one is interested in identifying a community structure that emphasizes differences in relationship type (e.g., coworkers, friends, or cohabitants) while still allowing some overlap. For example, a community of individuals that includes node replicates mostly from a work layer may also include node replicates in the bar layer (e.g., because some coworkers may often go to happy hour together after work). A small relaxation rate may mostly separate communities of different relationship types, but it would still allow mixing when some individuals have multiple relationship types with the same people (e.g., friends and coworkers).

Conversely, one may wish to use a larger relaxation rate for a multiplex network when the interactions on different layers are related to the same type of relationship (e.g., friendships, where individuals play sports, watch movies, and/or drink beer) and one is interested in identifying a community structure that emphasizes different social circles while still allowing some individuals to be part of multiple groups. For example, an individual may have a primary friend group that always watches movies and drinks together, but this individual may also be much more competitive than the others and plays in a completely different sports league (and thus is also part of a friendship circle of competitive athletes). A large relaxation rate in this scenario may allow communities to easily span multiple layers (as social circles usually engage in multiple interaction types), but it would still allow mixing when some individuals are involved in multiple social circles.

In the main text of the paper, we used a relaxation rate of 0.3 in our analysis of dolphin community structure (Section 2.2.1). We chose this value because it was the smallest relaxation rate that yielded qualitatively different structures when using multilayer InfoMap versus detecting communities as independent monolayer networks using InfoMap (Figures S3 and S4). For example, the former includes a community with node replicates from all layers.

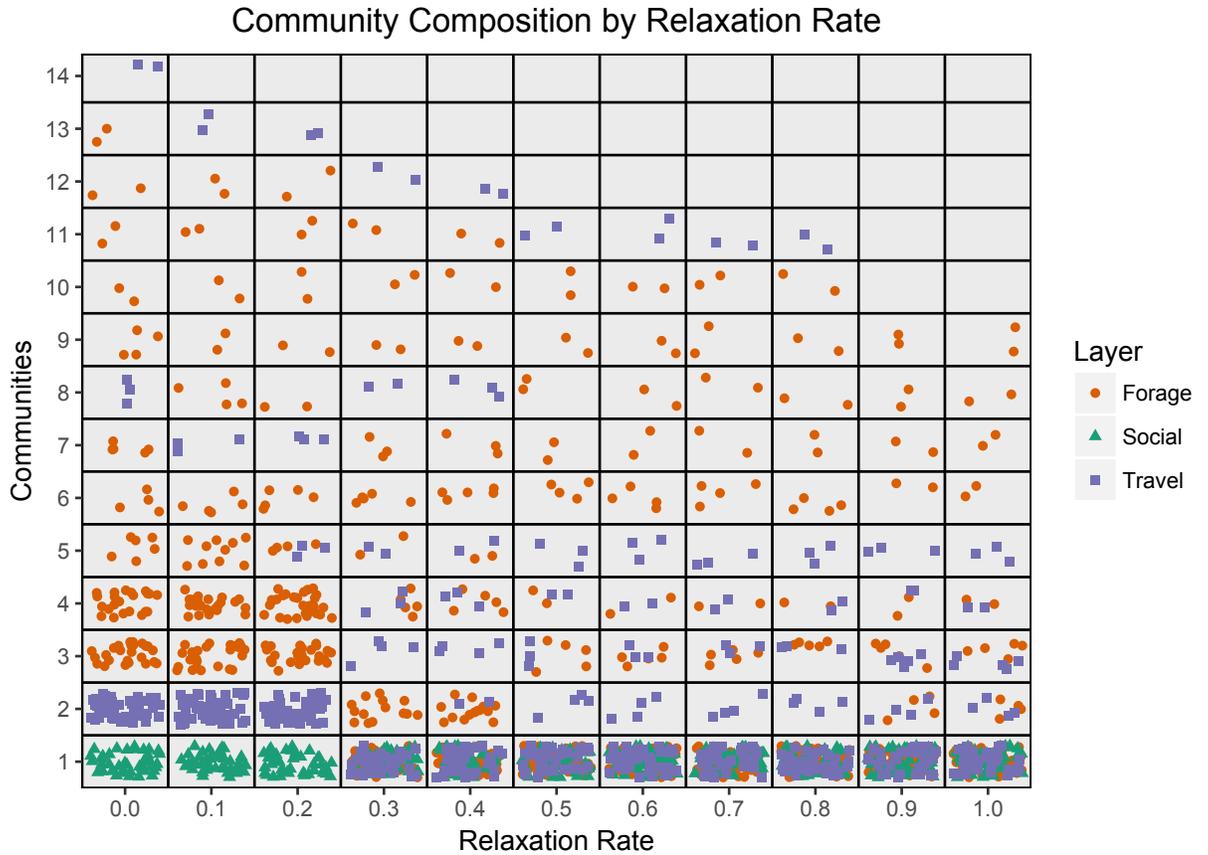

**Figure S3**: We investigate the number and composition of communities (on the vertical axis) across relaxation rates (on the horizontal axis) from 0 to 1, in increments of 0.1, for detecting communities using the multilayer InfoMap algorithm of De Domenico et al. (2015). We use the MuxViz implementation of the algorithm. Each point indicates an individual in a particular behavioral situation, which we designate by the colors of the points. Thus, in each column, an individual can appear up to three times (once in each color). Each cell in the grid represents a community, such that a certain cell includes all of the individuals in a particular community for a given relaxation rate. The number of communities and their composition (of nodes from the same or different layers) is different for different relaxation rates.

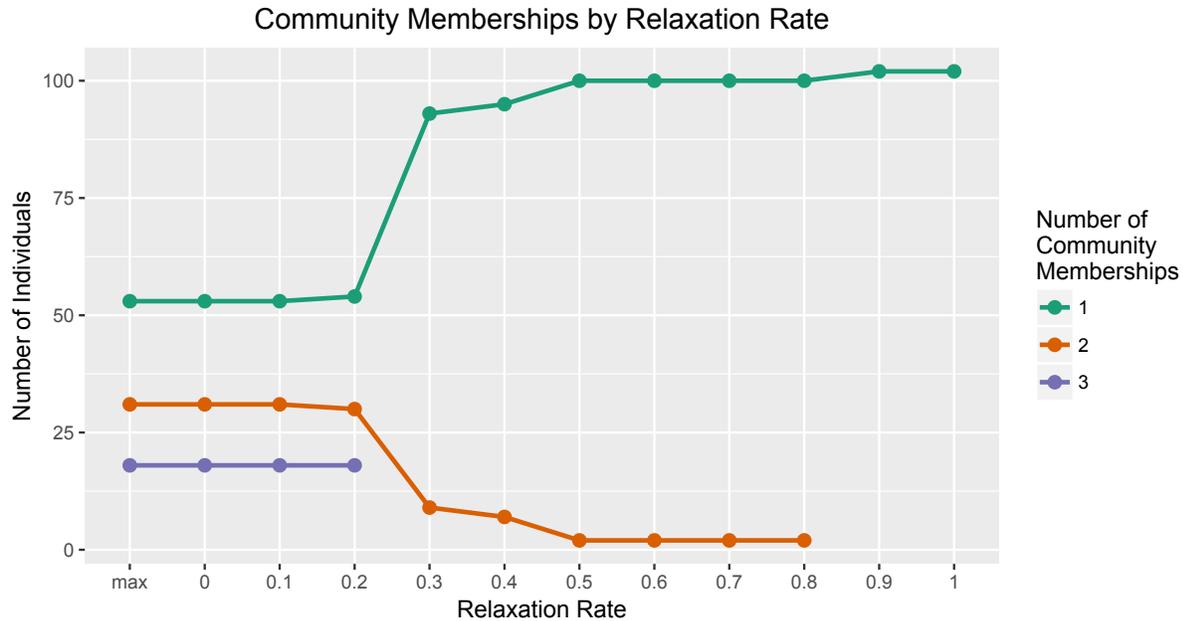

**Figure S4:** The number of communities of which individuals are members varies with the relaxation rate (horizontal axis). The maximum represents the number of layers in which an individual is present, so it is also the number of possible communities of which it can be a member. The number of individuals (vertical axis) who are in only one community (green) increases with relaxation rate. The number of individuals who are members of two communities (orange) decreases with relaxation rate. For relaxation rates of 0.3 and above, we no longer observe individuals who are members in three different communities (purple). Due to this qualitative difference in the number of communities in which individuals can be members between relaxation rates of 0.2 and 0.3, we choose 0.3 as the relaxation rate to implement in the main text.